\documentclass[lettersize,journal]{IEEEtran}
\usepackage{amsmath,amsfonts,amsthm,amssymb}
\usepackage{algorithmic}
\usepackage{algorithm}
\usepackage{array}
\usepackage[caption=false,font=footnotesize,labelfont=rm,textfont=rm]{subfig}
\usepackage{textcomp}
\usepackage{stfloats}
\usepackage{url}
\usepackage{verbatim}
\usepackage{graphicx}
\usepackage{cite}
\usepackage{multirow} 
\usepackage{xcolor}
\usepackage{bm}

\usepackage[hidelinks,bookmarks=false]{hyperref}

\def\R{\mathbb{R}}

\def\b{\mathbf{b}}
\def\x{\mathbf{x}}

\def\T{\mathsf{T}}    
\def\btheta{{\bm{\theta}}}
\def\bxi{{\bm{\xi}}}
\def\bXi{{\bm{\Xi}}}
\def\x{{\bf x}}

\def\f{{\bf f}}

\def\A{{\bf A}}

\def\Q{{\bf Q}}
\def\S{{\bf S}}

\def\0{{\bf 0}}

\def\0{{\bf 0}}

\def\Rn{{\mathbb{R}}}
\def\L{{\cal L}}    
\def\Ls{{\mathfrak{L}}}    
\def\bXi{{\bm{\Xi}}}

\theoremstyle{definition}
\newtheorem{defn}{Definition}

\begin{document}

\title{Physics-inspired Neural Networks for Parameter Learning of Adaptive Cruise Control Systems}

\author{Theocharis Apostolakis and Konstantinos Ampountolas,~\IEEEmembership{Member,~IEEE}
\thanks{Copyright \copyright~2024 IEEE. Personal use of this material is permitted. However, permission to use this material for any other purposes must be obtained from the IEEE by sending a request to pubs-permissions@ieee.org.}
\thanks{The publication of the article in OA mode was financially supported by HEAL-Link.}
\thanks{This work was supported by the Center of Research, Innovation and Excellence  (CRIE) of the University of Thessaly, Greece, under the project ``campaigningACC".}
\thanks{Th.~Apostolakis and K.~Ampountolas are with the Automatic Control and Autonomous Systems Laboratory, Department of Mechanical Engineering, University of Thessaly,  38334 Volos, Greece (e-mail: tapostolakis@uth.gr, k.ampountolas@uth.gr).}
\thanks{This paper includes supplementary  material available at \url{https://github.com/TheocharisAp/ACC_Identification_PINNs}, provided by the authors.}
}

\markboth{ IEEE Transactions on Vehicular Technology,~Vol.~XX, No.~Y, Month~2024}{Apostolakis and Ampountolas: Physics-inspired Neural Networks for Parameter Learning of ACC Systems}

\IEEEpubid{0000--0000/00\$00.00~\copyright~2024 IEEE}

\maketitle

\begin{abstract}
This paper proposes and develops a \emph{physics-inspired neural network} (PiNN) for learning the parameters of commercially implemented adaptive cruise control (ACC) systems in automotive industry. To emulate the core functionality of stock ACC systems, which have proprietary control logic and undisclosed parameters, the constant time-headway policy (CTHP) is adopted. Leveraging the multi-layer artificial neural networks as \emph{universal approximators}, the developed PiNN serves as a surrogate model for the longitudinal dynamics of ACC-engaged vehicles, efficiently learning the unknown parameters of the CTHP. The PiNNs allow the integration of physical laws directly into the learning process. The ability of the PiNN to infer the unknown ACC parameters is meticulously assessed using both synthetic and high-fidelity empirical data of space-gap and relative velocity involving ACC-engaged vehicles in platoon formation. The results have demonstrated the superior predictive ability of  the proposed PiNN in learning the unknown  design parameters of stock ACC systems from different car manufacturers. The set of ACC model parameters obtained from the PiNN revealed that the stock ACC systems of the considered vehicles in three experimental campaigns are neither  $\L_2$ nor $\L_\infty$  string stable.
\end{abstract}


\section{Introduction}\label{sec:Introduction}

\IEEEPARstart{T}{he} growing acceptance of partially automated vehicles on public roads, up to SAE Level 2 of driving automation \cite{SAE:2021}, has given rise to new traffic conditions. These conditions have raised concerns among the scientific community regarding their impact on traffic flow and capacity, necessitating a deeper understanding of the fundamental principles underlying these vehicles. 

Adaptive cruise control (ACC) \cite{Ioannou:1993} is an advanced driver-assistance system (ADAS) \cite{SAE:2021} that automatically adjusts the vehicle's speed to maintain a safe pre-defined distance from the vehicle in front or to reach the user-specified speed by accelerating or decelerating. ADAS has long been a part of automotive equipment, available as optional or standard features,  providing additional assistance to the driver by controlling the vehicle's longitudinal movement. It achieves this by monitoring the surrounding environment with several onboard sensors such as radar, lidar, and others.

The design of ACC remains publicly unknown, making ACC systems not yet fully understood. One of the key aspects behind ACC's design is the spacing policy adopted by ACC manufacturers, along with its unknown design parameters. The spacing policy specifies the pre-defined desired (time or space) distance between an ACC-engaged vehicle and the vehicle in front. Among the various spacing control strategies developed in the literature, the constant time headway policy (CTHP) \cite{Ioannou:1993} stands out as one of the most remarkable \cite{Swaroop:1999, Kanellakopoulos:1998}. Despite its simplicity, the CTHP is capable  of accurately reproducing the dynamics and driving behavior of ACC-engaged vehicles in platoons, as evidenced by several field trials, see e.g., \cite{Milanes:2014, Gunter:2021, Shladover:2014}.      

This paper introduces physics-inspired neural networks (PiNNs) for  learning the parameters of commercially implemented ACC systems using empirical observations of space-gap and relative velocity. The constant time-headway policy, speculated to be implemented in stock ACC systems of various makes \cite{Milanes:2014}, is adopted to emulate the longitudinal motion of ACC-engaged vehicles flocking in homogeneous platoons. The pursued PiNN is a deep learning data-driven approach based on the physical model of the CTHP, derived from first physical principles (double integrator), and control theory (PID-like control).

\IEEEpubidadjcol

The contributions of this work are threefold: 
\begin{itemize}
    \item The development, for the first time in the relevant literature, of PiNNs for learning the parameters of commercial ACC systems across various vehicle makes. The PiNNs allow the integration of physical laws directly into the
learning process.
    \item The inference of unknown ACC parameters (inverse optimization) using the proposed PiNN approach on both synthetic and empirical car-following data from several experimental campaigns, which include diverse driving conditions and behaviors, showcasing remarkable predictive accuracy.
    \item The rigorous assessment of ACC-engaged vehicles in platoon formation  through a thorough quantitative and qualitative analysis including the $\L_2$ and $\L_\infty$ string stability criteria.
    String unstable platoons in the presence of disturbancies are intruding since their dynamics  can lead to phantom traffic shockwaves and thus to traffic congestion and poor system throughput.  
\end{itemize}

The rest of the paper is structured as follows. Section \ref{sec:Literature} reviews the relevant literature. Section \ref{sec:Theoretical_Background} presents  the constant time-headway policy (CTHP). It also briefly presents the  $\L_2$ and $\L_\infty$  criteria for string stability in platoons. Section \ref{sec:PINNs}  introduces the proposed PiNN for parameter learning. Section \ref{sec:Application}  illustrates the application of the PiNN to both synthetic and empirical observations obtained from three real-life campaigns. It also presents the inferred parameters of stock ACC systems for various makes and discuss their string stability condition. Section \ref{sec:Conclusions} discusses the potential of this work. 

\emph{Notation:} The field of real and complex numbers are denoted by $\Rn$ and $\mathbb{C}$, respectively. The imaginary unit is denoted by $j$, where $j := \sqrt{-1}$. The space of Lebesgue measurable functions $f: \Rn \to \Rn$ such that $t \to |f (t)|^\mathsf{p}$ is integrable over $\Rn$ is denoted by ${\cal L}_\mathsf{p}$, here $\mathsf{p} =   2$ or $\mathsf{p} = \infty$ is used to discuss string stability. For $\mathsf{p} = \infty$ no integration is used, and instead, the norm on ${\cal L}_\infty$ is given by the essential supremum. Given a transfer function $H(j\omega)$, $\omega\in\Rn$, of a single-input single-output (SISO) system, the ${\cal H}_\infty$ norm of the system is defined as $\| H(j\omega) \|_{{\cal H}_\infty} = \sup_{\omega\in\Rn} |H(j\omega)|$.

\section{Relevant Literature}\label{sec:Literature}

The spacing policy of ACC systems is one of the main aspects constituting these vehicles, with a direct impact on traffic flow characteristics and driver's safety. The most commonly used is the constant time headway policy (CTHP) adopted in this paper, which suggests that the predefined distance between two consecutive vehicles is a linear function of the controlled vehicle's (ACC follower) velocity \cite{Ioannou:1993}. In addition, the constant spacing policy (CSP) suggests that the desired inter-vehicle spacing is independent of the ACC ego vehicle's longitudinal velocity \cite{Swaroop:1999}. Finally, the nonlinear variable time headway policy (VTHP) proposes a time headway which varies with the relative speed between the two adjacent vehicles \cite{Kanellakopoulos:1998}. A comparison of the aforementioned spacing and time headway policies can be found in \cite{Swaroop:1994}.

Among others, a major issue concerning the scientific community for decades, is whether ACC systems are string stable inside a platoon \cite{Swaroop:1996, Liang:1999, Besselink:2017, Ploeg:2014}. String stability is defined as the elimination of propagating disturbances upstream the platoon. In the literature it has been repeatedly reported that these systems tend to be string unstable under the presence of disturbances \cite{Milanes:2014,Knoop:2019, Makridis:2020, Makridis:2021, Gunter:2020, Wang:2021, Gunter:2021}. To better understand such phenomena, a broader study of ACC's principles is required.

Employing these type of control policies on ACC-equipped vehicles would entail string stable platoons when such vehicles are involving. However, it is well known that constant spacing policies (CSP) lead to string unstable platoons \cite{Seiler:2004, Jovanovic:2005, Middleton:2010}, while constant time headway policies (CTHP) are shown to be either string stable \cite{Swaroop:1996, Liang:1999, Ioannou:1993, Rajamani:2002} or unstable \cite{Milanes:2014, Knoop:2019, Makridis:2020, Makridis:2021, Gunter:2020, Wang:2021, Gunter:2021} based on experimental or simulation studies in the literature. Consequently, string stability is a key property to achieve smooth traffic flows, guarantying the absence of potential disturbances propagating upstream a platoon of vehicles and hence, avoiding the emergence of traffic jams, slinky effects and unsafe gaps.

Embedding these policies into simplified vehicle models to reproduce or imitate ACC vehicle's dynamics and their realistic effects on traffic flows, implies the awareness of realistic ACC parameter values. Consequently, ACC system's parameter identification is an important step towards its better understanding. Several studies used either online or offline optimization procedures to estimate the unknown system dynamics and parameters using synthetic or empirical data  \cite{Milanes:2014, Gunter:2020, Punzo:2005, Kesting:2008}. 

A dual unscented Kalman filter (DUKF) was proposed for the nonlinear joint state and CTHP parameter identification of commercially implemented ACC systems, using empirical data from a real-life car-following experimental campaign \cite{Ampountolas:2023}. Another study, developed two online methods to provide real-time system identification of ACC-equipped vehicles: i) Recursive least squares (RLS); and ii) Particle filtering (PF), on both synthetic and empirical data \cite{Wang:2021}. In addition, a Gaussian process-based model was designed to learn the personalized driving behavior based on both synthetic and naturalistic data, to design an ACC system suitable to driver's preferences \cite{Wang:2022}. 

However, the parameter identification problem of automated vehicles using empirical observations is challenging, since the underlying optimization problem is non-convex and it might be ill-conditioned under equilibrium driving conditions (i.e., where acceleration and space-gap reduce to zero), see \cite{Wang:2021,Ampountolas:2023}. In the latter case, the ACC system parameters cannot be uniquely identified, given input and output observations from the platoon, since the problem lacks both linear and nonlinear observability \cite{Ampountolas:2023}. 

A recently developed framework for inverse optimization is the so-called physics-inspired (or physics-informed) neural networks (PiNNs) \cite{Lagaris:1998, Raissi:2019, Karniadakis:2021}. PiNNs are semi-supervised artificial neural networks of dynamical systems governed by ordinary or partial differential equations (ODEs or PDEs) and observed data. A PiNN consists of two ingredients: (a) An artificial neural network representing a {\it physics-uninformed} surrogate predictor (parameterized by weights and biases); and (b) a residual network representing a {\it physics-and-data-informed} set of ODEs or PDEs for regularization. PiNN's implementation expands into solving both forward (\emph{data-driven solution}) and inverse (\emph{data-driven discovery}) non-linear problems, with various applications in engineering\cite{Haghighat:2021, Huang:2023, Mao:2020} and transportation \cite{Shi:2021, Huang:2022}. 

The superiority of physics-inspired neural networks (PiNNs) over traditional neural networks (NNs) in physics-based problems lies in the integration of physical laws directly into the learning process. These laws act as a means of regularization, constraining the space of feasible solutions. This integration enhances models' accuracy and generalizability, making them particularly effective for complex dynamical systems where data may be sparse or where data acquisition may be highly expensive \cite{Raissi:2019}.

This research contributes to the state-of-the-art by introducing PiNNs for  learning the parameter of commercially implemented ACC systems using empirical observations of space-gap and relative velocity. This problem is appertained to the second class of problems (data-driven inverse optimization problems) integrating the knowledge from both empirical observations and ACC system's  dynamics. The proposed PiNN is based on the physical model of the longitudinal motion of ACC-engaged vehicles, derived from first physical principles, and a PID-like control law for emulating the CTHP.   

\section{Modeling Adaptive Cruise Control Systems}\label{sec:Theoretical_Background}

In the sequel, the constant time-headway policy, embedded in a simplified dynamic vehicle model, is considered to imitate the longitudinal motion of ACC-engaged vehicles flocking in homogeneous platoons. The vehicle dynamics of the ACC ego vehicles are described by a double integrator, while acceleration is governed by the CTHP. Despite its simplicity, this model is able to reproduce the dynamics and driving behavior of ACC-engaged vehicles in platoons as shown in field trials \cite{Milanes:2014, Gunter:2021, Shladover:2014}.      

\subsection{Longitudinal Model and Constant Time-Headway Policy}

Consider a platoon of $M$ homogeneous ACC-engaged ego vehicles, where  vehicle $i$ (follower) follows vehicle $i - 1$ (leader). The leader of the platoon is indexed by $i = 0$ and might be affiliated to a human-driven vehicle (HDV). The longitudinal dynamics of the ACC equipped vehicles, while acceleration is governed by the CTHP (a PID-like control law), can be described by the following system of ordinary differential equations (ODEs) \cite{Liang:1999, Ioannou:1993}: 
\begin{align}\label{eq:diff_1}
	\dot{p}_i(t) &= \Delta v_i(t), \quad i = 1, 2, \ldots, M,\\ \label{eq:diff_2}
	\dot{v}_i(t) &= \alpha_i \bigl[p_i(t) - \tau_i v_i(t)\bigl] + \beta_i \Delta v_i(t),
\end{align}
where $p_i(t)$ [m] is the space-gap between two vehicles $i$ and $i-1$, i.e., the distance between the follower's front bumper and the leader's rear bumper; and $\Delta v_i(t) = v_{i-1}(t) - v_i(t)$ is the relative velocity between the vehicle $i$ and the vehicle $i-1$ in a platoon. In control law \eqref{eq:diff_2}, $\tau_i$ [s] represents the constant time headway the ACC-engaged vehicle $i$ strives to maintain with its leader $i-1$, which in turn indicates the desired space-gap, $\tau v_i(t)$, between the two vehicles. The two non-negative gains, $\alpha_i$ [1/s$^2$] and $\beta_i$ [1/s], control the trade-off between the space-gap difference,  $p_i(t) - \tau v_i(t)$, and the relative velocity $\Delta v_i$. Finally, parameter $\tau_i$ [s] can also be considered as the time-gap under equilibrium driving conditions: 
\begin{align}\label{eq:equil}
	v_{i-1}(t) -v_i(t) = 0 \, \, \text{ and }\,\, p_i(t) - \tau_i v_i(t) = 0, \,\, \forall \, i.
\end{align}

The CTHP parameters $\alpha_i$, $\beta_i$, and $\tau_i$ that characterize the ACC equipped vehicles are constant but unknown. Thus, the model \eqref{eq:diff_1}--\eqref{eq:diff_2} can be re-written in compact form as, 
\begin{equation}\label{eq:platoon}
\dot{\bxi}_i(t) = \f[\bxi_i(t), v_{i-1}(t), \bm{\varpi}_i], \quad i = 1, 2, \ldots, M,
\end{equation}
where $\bxi_i(t) = [p_i(t)\,\,\, v_i(t)]^\T$ is the state vector,  $\bm{\varpi}_i = [\alpha_i\,\, \beta_i \,\, \tau_i]^\T$ is the parameters vector (to be learned), and  $\f = [f_1\,\,\, f_2]^\T$ is a vector function that reflects the right-hand side of  \eqref{eq:diff_1}--\eqref{eq:diff_2}. Note that, despite the homogeneity assumption above for the vehicle characteristics, a different set of parameters $\bm{\varpi}_i$ is considered here for each vehicle within the platoon. The assumption of the same set of parameters $\bm{\varpi}_i \equiv \bm{\varpi}$   for each ego vehicle $i$ in the platoon is a special case and is also considered in Section \ref{sec:Application}. 

Finally, the CTHP model of each ACC-engaged vehicle in the platoon must satisfy the following \emph{rational driving constraints} \cite{Wilson:2011}: 
\begin{equation}
	\frac{\partial f_2}{\partial p_i} = \alpha_i \ge 0,
	\frac{\partial f_2}{\partial \Delta v_i} = \beta_i \ge 0,
	\frac{\partial f_2}{\partial v_i} = -\alpha_i \tau_i \le 0, \, \forall \, i.
\end{equation}

\subsection{String Stability and Parameter Identification}

A fundamental aspect of ACC systems in platoon formation is the {\it string stability} in the presence of unknown but bounded disturbances. String stability characterizes the ability of the CTHP (or other control policy for ACC equipped vehicles) to mitigate the potential amplification of random perturbations through a platoon of vehicles. 

The following definition characterizes the lack of 
upstream amplification of random perturbations through a platoon of vehicles.
\begin{defn}[Strict String Stability]\label{defn:String_Stability}
Consider a string of $M$ vehicles in platoon formation. This system is strict string stable if and only if,
\begin{equation}\label{def:String_stab}
\|\chi_i(t)\|_{\L_\mathsf{p}} \le \|\chi_{i-1}(t)\|_{\L_\mathsf{p}}, \quad \forall\, t \ge 0, \,\, i =1, \ldots, M,
\end{equation}
where $\chi_i(t)$ can be any signal of interest for vehicle $i$ (e.g., spacing, velocity, or acceleration). The  input signal $\chi_0(t) \in \L_\mathsf{p}$ is given and corresponds to the lead vehicle of the platoon, while $\chi_i(0) =0$ for $i = 1, \ldots, M$. 
\end{defn}

Considering input-output stability for a platoon of $M$ vehicles, the transfer function (of two consecutive vehicles) from input velocity $V_{i-1}(s)$ to  output velocity $V_{i}(s)$ is defined by,
\begin{equation}\label{eq:tf}
    V_{i}(s) =H_i(s) V_{i-1}(s), \quad i = 1, 2, \ldots, M, 
\end{equation}
with $V(s)$, $s \in \mathbb{C}$, denoting the Laplace transform of $v(t)$, $t\ge 0$. Using \eqref{eq:tf}, the speed-to-speed transfer function of the CTHP model \eqref{eq:diff_1}--\eqref{eq:diff_2} for vehicle $i$ is given by, 
\begin{equation}\label{eq:tf_CTHP}
	H_i(s) = \frac{\beta_i s + \alpha_i}{s^2 + (\alpha_i \tau_i +  \beta_i)s + \alpha_i}, \quad i = 1, \ldots, M.
\end{equation}

Given the above transfer function of the CTHP, the following conditions for string stability in the sense of $\mathcal{L}_2$ and $\mathcal{L}_\infty$ norms are well-established (see e.g.,  \cite{Wilson:2011, Monteil:2019}).
\subsubsection{$\mathcal{L}_2$ Strict String Stability}
For the $\L_2$ space, the following condition holds \cite{Boyd:1991}:
\begin{equation}\label{eq:HinfH2relation}
\| H_i(j\omega) \|_{{\cal H}_\infty} = \max_{v_{i-1} \neq 0}\frac{\|v_{i}(t)\|_{\L_2}}{\|v_{i-1}(t)\|_{\L_2}}, \quad i = 1, \ldots, M, 
\end{equation}
where $H_i(j\omega)$ is the transfer function evaluated at $s = j\omega$ over the frequency $\omega \in \Rn$ (i.e., along the imaginary axis). According to \eqref{eq:HinfH2relation} the ${\cal H}_\infty$ norm is induced by the $\L_2$ norms of the input and output energy signals. Combining \eqref{eq:HinfH2relation} and Definition \ref{defn:String_Stability} for the $\L_2$ norm yields the following condition for strict string stability (see also Notation in Section \ref{sec:Literature}):
\begin{equation}\label{eq:L2_f}
    | H_i(j\omega) | =  \sqrt{\frac{\alpha^2_i +\beta^2_i \omega^2_i}{(\alpha_i-\omega_i^2)^2+\omega_i^2(\alpha_i \tau_i +  \beta_i)^2}} < 1, 
\end{equation}
for all $\omega \ge 0$ and $i = 1, \ldots, M$. This condition imposes the disturbance effect to decrease for long platoons rather than just being bounded (i.e., the disturbance will ultimately vanish).

Condition \eqref{eq:L2_f} leads to the following inequality  that the CTHP model parameters $\alpha_i$, $\beta_i$ and $\tau_i$ must satisfy,
\begin{equation}\label{eq:L2c}
\alpha_i^2 \tau_i^2 + 2\alpha_i \beta_i \tau_i - 2\alpha_i > 0,\, \forall\, i = 1, \ldots, M.
\end{equation}
As can be seen, as $\tau_i$ approaches $\infty$ the system is $\mathcal{L}_2$ strict string stable for all non-negative gains $\alpha_i$ and $\beta_i$ of the CTHP, while as $\tau_i$ approaches zero the system becomes unstable. 
\subsubsection{$\mathcal{L}_\infty$ Strict String Stability}
The transfer functions \eqref{eq:tf_CTHP} have second-order dynamics and a zero on the real axis of the left half complex plane.  
For the $\L_\infty$ space, a necessary and sufficient condition for string stability is non-imaginary poles and negative zeros in \eqref{eq:tf_CTHP}, which leads to:

\begin{equation}\label{eq:Linfc}
(\alpha_i \tau_i +  \beta_i)^2 - 4\alpha_i > 0.
\end{equation}

Subtracting \eqref{eq:L2c} from \eqref{eq:Linfc} yields \cite{Monteil:2019}: 
\begin{equation}\label{eq:L2Linf}
\beta_i^2 >  2\alpha_i \;\; \Rightarrow\;\; \left(\text{$\Ls_\infty$ stability $\Leftrightarrow$ $\Ls_2$ stability}\right),
\end{equation}
suggesting that  $\mathcal{L}_2$ stability is stronger than the $\mathcal{\L}_\infty$ stability. This is realistic since even if the $\mathcal{L}_2$ energy of a signal is small, it may occasionally contain large peaks, provided the peaks (i.e., the $\mathcal{L}_\infty$ norm) are not too frequent and do not contain too much energy. 

A proper parameter identification scheme is essential for the assessment of commercially implemented ACC systems in terms of string stability. Section \ref{sec:PINNs} presents a physics-constrained and data-informed artificial neural network for the parameter learning problem of stock ACC systems using empirical data of space-gap and relative velocity. This is a deep learning data-driven approach subject to the physical model of the CTHP \eqref{eq:diff_1}--\eqref{eq:diff_2}, derived from first physical principles (double integrator), and control theory (a PID-like control law).

\section{Physics-Inspired and Data-Informed\\ Artificial  Neural Networks}\label{sec:PINNs}

\subsection{Architecture of Multi-layer Artificial Neural Networks}\label{sec:ArchPINN}

Consider a fully-connected feedforward artificial neural network (ANN), ${\cal N}^{L}(\x): \R^n \to \R^m$ of $L$ (or $L-1$ hidden) layers, with $N_{\ell}$ denoting the number of artificial neurons at layer  ${\ell} = 1, 2, \ldots, L-1$, while for the input and output layers $N_{0} = n$ and $N_{L} = m$ holds, respectively. Thus, the input layer has the dimension of the raw training data $\x\in\R^n$, while the dimension of the output layer is defined by the context. In feed-forward ANNs, each neuron at layer $\ell$ is equipped with a (user-defined) nonlinear activation function, $\sigma(\x): \R^n\to  \R^n$, to transform the weighted linear sum input of various neurons at layer $\ell-1$ into an output that is passed (neuron fires) on to the next (hidden or output) layer $\ell+1$. The most common activation functions include the logistic sigmoid $\sigma(x)=1/[1+\exp(-x)]$, hyperbolic tangent, $\sigma(x) = \tanh (x)$, and the rectified linear unit (so-called ReLU), $\sigma(x) = \max (x, 0)$. The weights and bias of the weighted sum input  at layer $\ell = 1, 2, \ldots, L$ are organized in the matrices $\A^{\ell} \in \R^{N_{\ell}\times N_{\ell-1}}$ and vectors $\b^{\ell} \in \R^{N_{\ell}}$, respectively; or in a single parameters vector $\btheta = \{\A^{\ell},\b^{\ell}\}_{1 \le \ell \le L}$. The architecture of the ANN can then be summarized as the following compositional function:
\begin{align*}
\text{\it Input layer:}\quad {\cal N}^{0}(\x) &= \x \in \R^n,\\
\text{\it Hidden layers:}\quad {\cal N}^{\ell} (\x) &= \sigma(\A^{\ell}{\cal N}^{\ell-1} (\x)+\b^{\ell}) \in \R^{N^{\ell}},\\ 
\text{\it Output layer:}\quad {\cal N}^{L}(\x) &= \A^{L}{\cal N}^{L-1} (\x)+\b^{L} \in \R^m,
\end{align*}
where $\ell = 1, 2, \ldots, L-1$ for the hidden layers. 

The ultimate goal is to optimize the weighting matrices $\A^{\ell}$ and bias vectors $\b^{\ell}$ connecting the neural network from the $\ell$-th to $(\ell+1)$-th  layer using labeled pairs of input (see \emph{Input layer}) and output (see \emph{Output layer}) data. This procedure, which is known as \emph{training} in the ANN nomenclature, involves \emph{automatic differentiation} \cite{Baydin:2018} over the compositional function ${\cal N}^{L}(\x)$ and nonlinear optimization over a high-dimensional space, and thus is computationally expensive. \emph{Stochastic gradient descent} (SGD) \cite{Robbins:1951,Polyak:1992} and \emph{back-propagation} \cite{Rumelhart:1986} are two important algorithms that can be used for the efficient training of such multi-layer ANN.

\subsection{Parameter Learning via Physics-inspired Neural Networks}

Consider the  parameterized ODEs  \eqref{eq:diff_1}--\eqref{eq:diff_2} that characterize the CTHP of ACC equipped vehicles. Assume that a solution, $\bxi_i(t)$ on the time domain $t \in [t_0, t_e] = \Upsilon$, of the ODEs exists, given some boundary conditions ${\cal B}_i(\bxi_i,t) =\0$ on $\partial \Upsilon$:
\begin{equation}\label{eq:platoonIP}
{\cal F}_i(\bxi_i, \dot{\bxi}_i, \bm{\varpi}_i, t) = \dot{\bxi}_i(t) - \f[\bxi_i(t), \bm{\varpi}_i] = \0,
\end{equation}
for all $i = 1, \ldots, M$, where ${\cal F}_i\in \Rn^2$ is a nonlinear operator (residual of \eqref{eq:platoon}) parameterized by $\bxi_i$, $\dot{\bxi}_i$, and $\bm{\varpi}_i$ (to be learned).

\begin{figure*}[tbp]
     \centering
         \includegraphics[width=.9\linewidth]{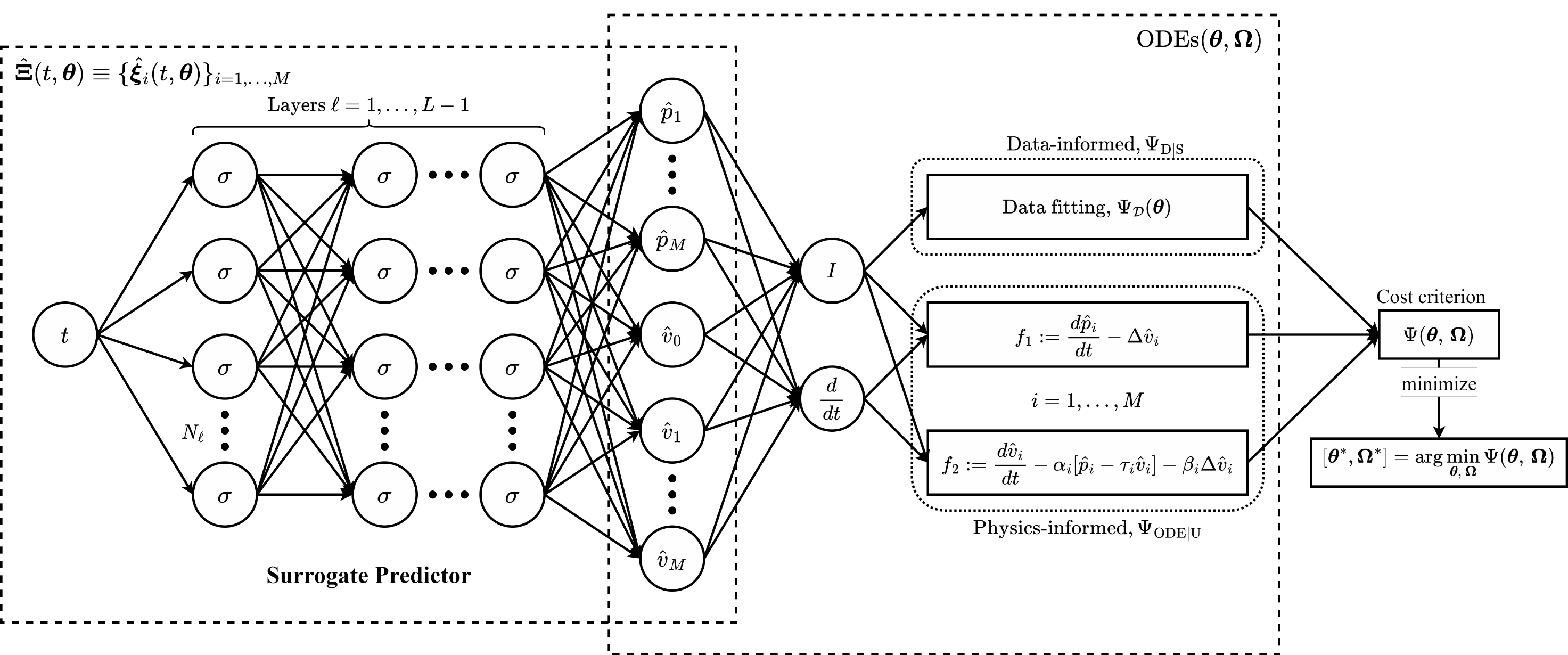}
         \caption{Architecture of the proposed physics-inspired and data-informed neural network for CTHP parameter learning. The neural network on the left represents the {\it physics-uninformed} surrogate predictor $\hat{\bXi}(t;\bm{\theta})= \{\hat{\bxi}_i(t;\bm{\theta})\}_{i=1, \ldots, M}$, while the right network depicts the {\it physics-inspired} residual and boundary conditions of the CTHP, and the data-informed cost function that penalizes deviation of the surrogate model predictions from the empirical data.}
    \label{fig:Neural_Net}
\end{figure*}

Multi-layer feedforward artificial neural networks are a class of \emph{universal approximators} \cite{Hornik:1989}, thus a neural network (see Fig.\ \ref{fig:Neural_Net}), $\hat{\bm{\Xi}} = \{\hat{\bxi}_i(t;\bm{\theta})\}_{i=1, \ldots, M}$, can be developed as a surrogate of the solution $\bxi_i(t)$, for all $i = 1, 2, \ldots, M$. Provided the availability of empirical data of space-gap and relative velocity, ${\cal D} = \{\bxi_i(t)\}_{t\in \Upsilon}$, the (output of the) neural network $\hat{\bm{\Xi}}$ (predictor) can be constrained to satisfy the physical model \eqref{eq:platoonIP} of the CTHP and its boundary conditions ${\cal B}_i(\bxi_i, t) =\0$ on $\partial \Upsilon$, for all $i = 1, 2, \ldots, M$. Additional internal conditions, ${\cal I}_i(\bxi_i,t) =\0$ on some $t \subset \Upsilon$,  can be also incorporated for readily solving the \emph{inverse parameter optimization problem}. In the inverse problem, the vector of CTHP parameters are to be learned using training data, ${\cal D}$, such that \eqref{eq:platoonIP} and boundary/internal conditions are satisfied. 

Concluding, a physics-inspired and data-informed neural network for each ACC vehicle $i$ on a platoon consists of (see Fig.\ \ref{fig:Neural_Net}): (a) the {\it physics-uninformed} surrogate predictor $\hat{\bxi}_i(t;\bm{\theta})$; (b) the {\it physics-informed} residual constraints ${\cal F}_i(\bxi_i, \dot{\bxi}_i, \bm{\varpi}_i, t)$ together with the boundary and internal conditions  ${\cal R}_i(\bxi_i, t) = \{{\cal B}_i(\bxi_i, t) \cup {\cal I}_i(\bxi_i,t)\}$; (c) a \emph{data-informed} fitting function that penalizes deviations of neural network's predictions from the empirical training data. Upon training, the neural network is calibrated to predict the entire solution of the system of ODEs \eqref{eq:platoonIP}, as well as the unknown CTHP parameters that impose the underlying dynamics in a platoon.

\subsection{Training of the Physics-inspired Neural Network}

Training of the neural network requires discretization or sampling of the continuous time domain on a sufficient small time interval, resulting to the set of pseudo-points (or induced points) ${\cal T} = \{t_1, t_2, \ldots,t_{|{\cal T}|}\} \in \Upsilon$, where $|{\cal T}|$ is the cardinality of set $\cal T$. The set ${\cal T}=\{{\cal T}_f,{\cal T}_b\}$ includes all points in the time domain ${\cal T}_f \subset \Upsilon$ and its boundary ${\cal T}_b \subset \partial \Upsilon$ where evaluation of the residual \eqref{eq:platoonIP} and boundary and internal conditions  ${\cal R}_i$ is necessary, given predictions of the surrogate model $\hat{\bxi}_i(t;\bm{\theta})$ and empirical data of space-gap and relative-velocity, ${\cal D}  = \{\bxi_i(t)\}_{t\in \cal T}$. Thus, the neural network takes as input the set ${\cal T}$ and provides predictions of $\hat{\bxi}_i= [\hat{p}_i \,\, \hat{v}_i]^\T$ (see the input and output layers, respectively, in Fig.\ \ref{fig:Neural_Net}), for all  $i = 1, 2, \ldots, M$. The leader's velocity, $v_0(t)$, can be also predicted by the NN, provided that $v_0(t)$ is observed using range sensors or other equipment, see $\hat{v}_0$ in Fig.~\ref{fig:Neural_Net}.

To efficiently train the surrogate predictor $\hat{\bm{\Xi}} = \{\hat{\bxi}_i(t;\bm{\theta})\}_{i=1, \ldots, M}$ and simultaneously satisfy the residual and boundary constraints the following cost criterion of two terms is considered: (a) a semi-unsupervised cost criterion, ${\Psi}_{\rm ODE|U}$,  for the residual \eqref{eq:platoonIP} and its boundary and internal conditions; (b) a supervised (data-driven) cost criterion, ${\Psi}_{\rm D | S}$, governed by measurements of $\bxi_i$ (empirical data of space-gap and relative-velocity) and the predictions $\hat{\bxi}_i$, for all $i=1, \ldots, M$. For the inverse optimization problem the cost criterion reads:
\begin{align}\notag
{\Psi}(\btheta , \bm{\Omega}) &= {\Psi}_{\rm ODE|U} + {\Psi}_{\rm D | S}\\ \label{eq:PINN_Loss}
		&=  \underbrace{{{\Psi}}_{\cal F}(\btheta , \bm{\Omega}) + {{\Psi}}_{\cal R}(\btheta , \bm{\Omega})}_{{\Psi}_{\rm ODE|U}} + \underbrace{{\Psi}_{\cal D}(\btheta)}_{{\Psi}_{\rm D | S}},
\end{align}
where $\bm{\Omega} = \big[\bm{\varpi}_1^\T\,\,\cdots\,\,\bm{\varpi}_M^\T\big]^\T$, and
\begin{align}
\Psi_{\cal F}(\btheta , \bm{\Omega}) &= \frac{1}{|{\cal T}_f|\times M}\sum_{i = 1}^M \sum_{t \in {\cal T}_f}\|{\cal F}_i(\hat{\bxi}_i, \dot{\hat{\bxi}}_i, \bm{\varpi}_i, t)\|^2_{\Q},
\end{align}
\begin{align}
\Psi_{\cal R}(\btheta , \bm{\Omega}) &= \frac{1}{|{\cal T}_b|\times M}\sum_{i = 1}^M \sum_{t \in {\cal T}_b} \| {\cal R}_i(\hat{\bxi}_i, \bm{\varpi}_i, t) \|^2_{\bf R},\\
{{\Psi}}_{\cal D}(\btheta) &=  \frac{1}{|{\cal T}|\times M}\sum_{i = 1}^M \sum_{t \in {\cal T}} \| \psi_i(\hat{\bxi}_i, \bxi_i, t)\|^2_{\S},
\end{align}
and $\psi$ is a function that penalizes deviations of the neural network surrogate approximation from the empirical training dataset ${\cal D}  = \{\bxi_i(t)\}_{t\in \cal T}$, e.g., for the mean square error (MSE), $\psi_i(\hat{\bxi}_i, \bxi_i, t) = \hat{\bxi}_i - \bxi_i$. The positive-definite weighting matrices $\Q$,  $\bf R$,  $\S$ are penalty terms for the unsupervised and supervised  cost functions, respectively. These matrices can be selected via a trial-and-error procedure to speed up convergence for a particular dataset and application.

The optimal vectors of the inverse problem, NN weights $\btheta$ and CTHP parameters $\bm{\Omega}$ for a platoon of vehicles, can be obtained by solving the following optimization problem during neural network's training:
\begin{equation}\label{eq:Inv_Opt}
	[\btheta^\ast, \bm{\Omega}^\ast]= \arg \min_{\btheta,\, \bm{\Omega}} {\Psi}(\btheta,\, \bm{\Omega}).
\end{equation}
This is a highly nonlinear optimization problem over a high-dimensional space, and thus is computationally expensive; but can be solved efficiently using (batch) SGD or quasi-Newton methods using Hessian information (e.g., Adam \cite{Kingma:2014} and L-BFGS \cite{Byrd:1995}).

\section{Application and Results}\label{sec:Application}

To demonstrate the effectiveness of the proposed data-driven approach to learn the parameters of the CTHP of ACC equipped vehicles, three physics-constrained and data-informed neural networks are developed and tested to both synthetic and empirical  data of space-gap and relative velocity. The first neural network is trained to predict the CTHP parameters for a car-following scenario with an ACC ego vehicle (follower) and a human-driven vehicle (leader), while two other neural networks are trained to predict the CTHP parameters for two different platoons of vehicles.

\subsection{Empirical Data Description}\label{sec:ED}

The empirical data are taken from three car-following experimental campaigns that took place in Autostrada A26 motorway in Ispra-Vicolungo (fleet of five vehicles in platoon formation) and Ispra-Casale  (car-following scenario with two vehicles) routes in 2019 and 2020, respectively, Italy; and AstaZero test track (five premium ACC equipped vehicles in platoon formation) in mid 2019, Sweden. In all campaigns, data acquisition was performed using high-accuracy on-board equipment (e.g., U-blox global navigation satellite system (GNSS) receivers) with a sampling frequency at 10 Hz (0.1 s). These data can be freely accessed through the OpenACC database \cite{Makridis:2021}. Table \ref{tab:New_Both}  provides information (make and model) on the vehicles involved in the  campaigns.

\subsection{Physics-inspired Neural Network Setup}\label{sec:PINN_Setup}

Fig.~\ref{fig:Neural_Net} depicts the neural network architecture for parameter learning of the unknown design parameters of the CTHP, $\bm{\Omega}$, for a platoon of $M$ vehicles. Each neuron of the neural network is equipped  with a non-linear hyperbolic tangent activation function, $\sigma(\cdot) = \tanh(\cdot)$.

The first neural network was developed and trained to predict the CTHP parameters for a particular ego vehicle $i$ in a car-following scenario (i.e., $M=1$) in Ispra-Casale. The network's architecture is composed of an input layer with one (1) neuron $t$ (the discretized time domain), three (3) hidden-layers with sixty (60) neurons each, and an output layer with three (3) neurons, the predictor  $\hat{\bm{\Xi}}(t;\bm{\theta}) = \hat{\bxi}_1(t;\bm{\theta})$ plus the leader's velocity $\hat{v}_0$, and hence, the output $(\hat{v}_0, \hat{\bxi}_1) \in  \R^3$ where $\hat{\bxi}_1= [\hat{p}_1\,\, \,\hat{v}_1]^\T$. Thus (see Section \ref{sec:ArchPINN} for notation), $L=4$, $N_{0} = 1$, $N_{\ell} =60$ for $\ell = 1, 2, 3$, and $N_4 = 3$. The weights vector of the neural network reads $\bm{\theta} = \{\A^{(1)}, \A^{(2)}, \A^{(3)}, \A^{(4)}, \b^{(1)}, \b^{(2)}, \b^{(3)}, \b^{(4)}\} \in \R^{7623}$.

Two additional neural networks with the same architecture ($L=4$, $N_{0} = 1$, $N_{\ell} =60$, and $\ell = 1, 2, 3$) were trained to predict the CTHP parameters of five vehicles in platoon formation for two different experimental campaigns, namely AstaZero and Ispra-Vicolungo. For the AstaZero dataset, $M=4$, and thus the output layer of the neural network consists of nine (9) neurons, the predictor $\hat{\bm{\Xi}} = \{\hat{\bxi}_i(t;\bm{\theta})\}_{i=1, \ldots, 4}$ (i.e., $(\hat{p}_i, \hat{v}_i)$ for $i = 1, \ldots, 4$) plus the leader’s velocity $\hat{v}_0$, i.e., $N_4 = 9$. For the Ispra-Vicolungo dataset, the first and the last vehicle in the platoon are human driven, thus $M=3$, and the output layer of the neural network consists of seven (7) neurons, the predictor $\hat{\bm{\Xi}} = \{\hat{\bxi}_i(t;\bm{\theta})\}_{i=1, \ldots, 3}$ (i.e., $(\hat{p}_i, \hat{v}_i)$ for $i = 1, \ldots, 3$) plus the leader’s velocity $\hat{v}_0$, i.e., $N_4 = 7$. The weights vectors of the two neural networks read: $\bm{\theta} = \{\A^{(1)}, \A^{(2)}, \A^{(3)}, \A^{(4)}, \b^{(1)}, \b^{(2)}, \b^{(3)}, \b^{(4)}\} \in \R^{7989}$ and $\bm{\theta} = \{\A^{(1)}, \A^{(2)}, \A^{(3)}, \A^{(4)}, \b^{(1)}, \b^{(2)}, \b^{(3)}, \b^{(4)}\} \in \R^{7867}$ for the AstaZero and Ispra-Vicolungo campaigns, respectively.

To train the neural networks, the learning rate is set to $0.001$, and the maximum number of iterations is set to 60,000. In each iteration, the input of the neural network is fed with a set of randomly sampled collocation (pseudo-) points within the time domain of $t = [0, 300]$ s. Data includes measurements at a frequency of 10 Hz (0.1 s), i.e., at \ ${\cal T} = \{t_1, t_2, \ldots,t_{|{\cal T}|=3001}\}$ (see input $t$ of the input layer in Fig.\ \ref{fig:Neural_Net}), and was divided into training and test sets. Thus, in each iteration, the whole training set, which includes points randomly sampled within the time domain, is used to train the PiNN (full-batch training). After the training phase, the test set is utilized to validate the PiNN's ability to generalize effectively to unseen data and to reconstruct the full range of dynamics within the examined time domain.

For the inverse optimization problem for a car-following scenario with two vehicles, $M = 1$ (the extension to $M>1$ is straightforward) the cost criterion \eqref{eq:PINN_Loss}, is considered with:
\begin{align*}
&\Psi_{\cal F}(\btheta , \bm{\omega}) = \frac{1}{|{\cal T}_f|}  \sum_{t \in {\cal T}_f} \left\{ \big[ \hat{\dot{p}}_1(t) - \hat{v}_0(t) + \hat{v}_1(t) \big]^2 \right. \\ 
                         &\quad\,\,\, + \left.  \big[ \hat{\dot{v}}_1(t) - \alpha [\hat{p}_1(t) - \tau \hat{v}_1(t)] - \beta [\hat{v}_0(t) - \hat{v}_1(t)]\big]^2 \right\},\\
&{{\Psi}}_{\cal D}(\btheta) =  \frac{1}{|{\cal T}|} \sum_{t \in \cal T} \left\{ \left[p_1(t) - \hat{p}_1(t)\right]^2 + \left[v_1(t) - \hat{v}_1(t)\right]^2 \right.\\
&\qquad\quad\quad\qquad\qquad\qquad\qquad\qquad\,\,\,\, \left.  + \left[v_0(t) - \hat{v}_0(t)\right]^2 \right\},
\end{align*}
where $v_0$ and $v_1$ are the velocities of the leader (ACC vehicle or HDV) and follower (ACC ego vehicle), respectively; and $p_1$ is the space-gap between the two vehicles. Note that the term $\Psi_{\cal R}$ is absent above, since   $|{\cal T}_b| = |{\cal T}|$ and thus the internal and boundary conditions ${\cal R}$ are included in $\Psi_{\cal F}$ and $\Psi_{\cal D}$. For the inverse optimization problem  \eqref{eq:Inv_Opt} a hybrid strategy of using Adam (for some thousand of iterations at the beginning of the training) and L-BFGS (for the rest of the training) could be employed to speed-up the convergence. 

The described neural network architecture (among others tested) is found to work well for the CTHP parameter learning problem and, thus, it has been employed for both synthetic  and empirical data obtained from campaigns in the field (see Section \ref{sec:ED}). Training of the neural network takes about 7 CPU-minutes for the Ispra-Casale experiment and 30 CPU-minutes for the two other campaigns with platoons of $M=4$ and $M=3$ in a computer with an 11th Generation Intel(R) Core(TM) i7-11700K @ 3.60GHz with 8 cores on Windows 10 Pro 64-bit. Parameter convergence is achieved in around 20,000 iterations for all tests on synthetic and empirical data.

\begin{figure*}[tbp] 
    \centering
  \subfloat[Space-gap: Synthetic data, solid line; PiNN prediction, dashed line.]{%
       \includegraphics[width=.5\linewidth]{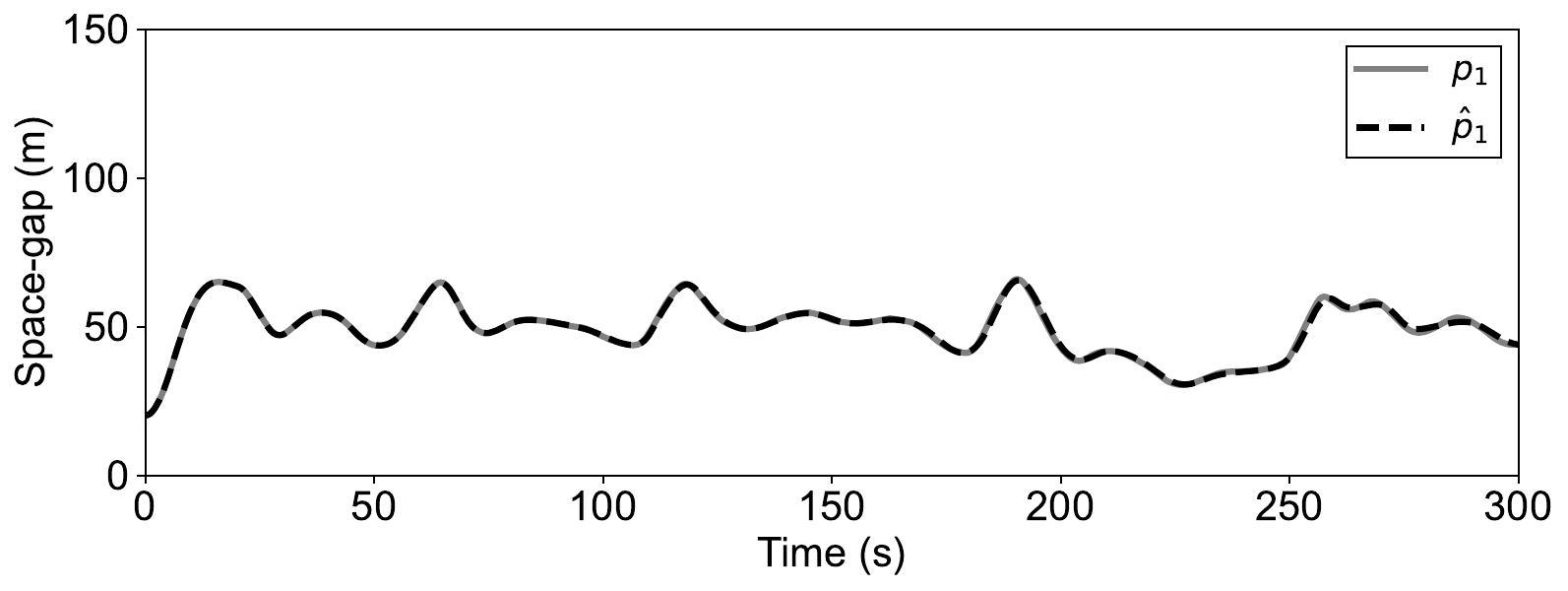}}
    \hfill 
  \subfloat[Velocity: Synthetic data, solid line; PiNN prediction, dashed line.]{%
        \includegraphics[width=0.5\linewidth]{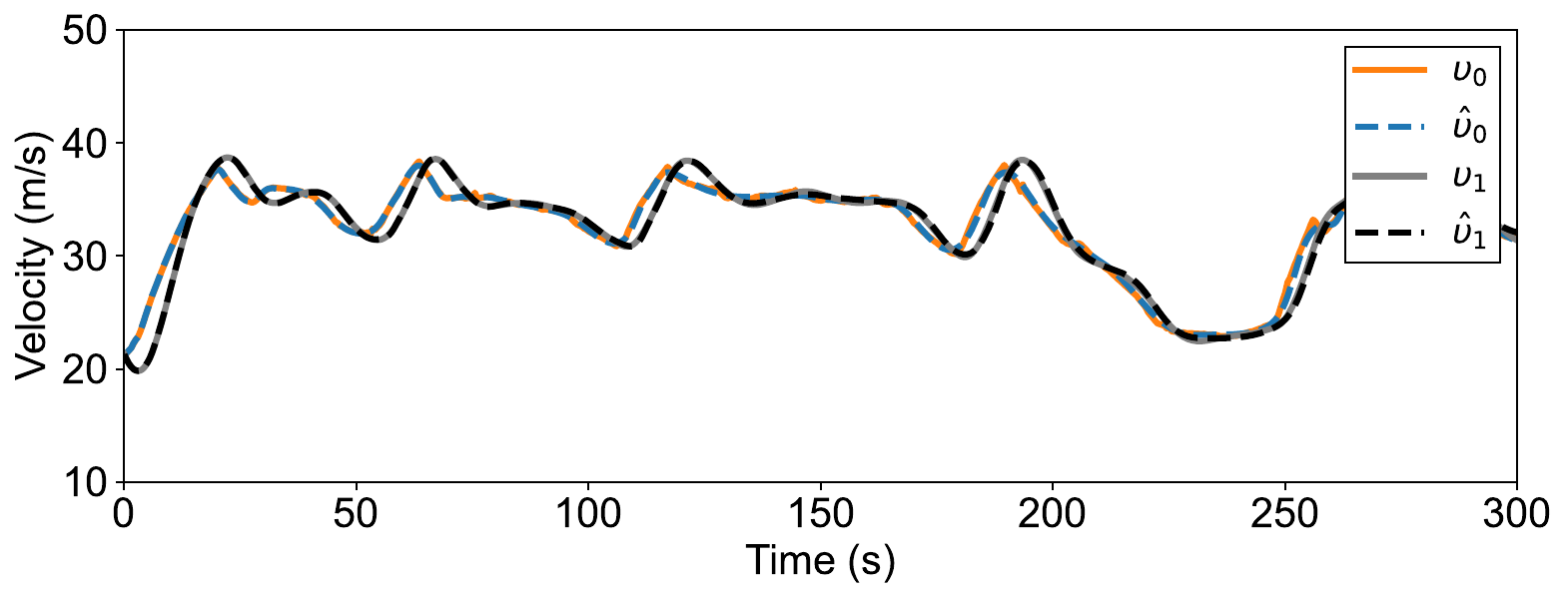}}
      \caption{Space-gap and velocity trajectories for the synthetic dataset.}
  \label{Synthetic data} 
\end{figure*}

\subsection{CTHP Parameter Learning on Synthetic Data} 

Initially, synthetic data with known CTHP parameters, $\bm{\omega}$, were generated and used to validate the proposed  parameter learning approach. To this end, consider a car-following-scenario with two vehicles (i.e., $M=1$), a leader (HDV) and a follower (ACC ego vehicle), a known set of design parameters $\bm{\varpi}^\ast = [\alpha^\ast\,\, \beta^\ast \,\, \tau^\ast]^\T = [0.08\,\,\, 0.12\,\,\, 1.5]^\T$, and a pre-defined leader's velocity profile $v_0(t)$ taken from the Ispra-Casale campaign (see the solid orange line in Fig.~\ref{Synthetic data}b). Note that we omit the index and use $\bm{\varpi}^\ast$ instead of $\bm{\varpi}_{1}^\ast$ for simplicity. Then, synthetic data of $p_1(t)$ and $v_1(t)$, for $t>0$, were generated from the system of ODEs \eqref{eq:diff_1}--\eqref{eq:diff_2}, using the a priori known $\bm{\varpi}^\ast$ and the initial conditions $\bxi_{1}(0) = [p_1(0) \,\,\, v_1(0)]^\T  = [20.3 \,\,\, 21.3]^\T$ (in m and m/s, respectively).
The data was generated for 300 s with a frequency of 10 Hz (0.1 s).

\begin{figure}[tbp] 
    \centering
  \subfloat[Synthetic data\label{Synthetic_param}]{%
       \includegraphics[width=0.49\linewidth]{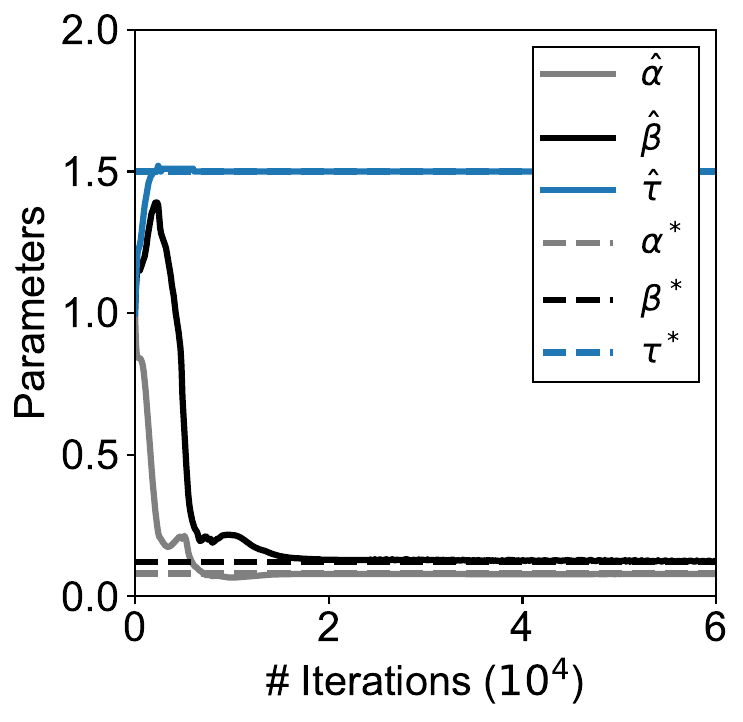}}
    \hfill
  \subfloat[Ispra-Casale (Exp.~\#1)\label{Ispra-Casale_param}]{%
        \includegraphics[width=0.49\linewidth]{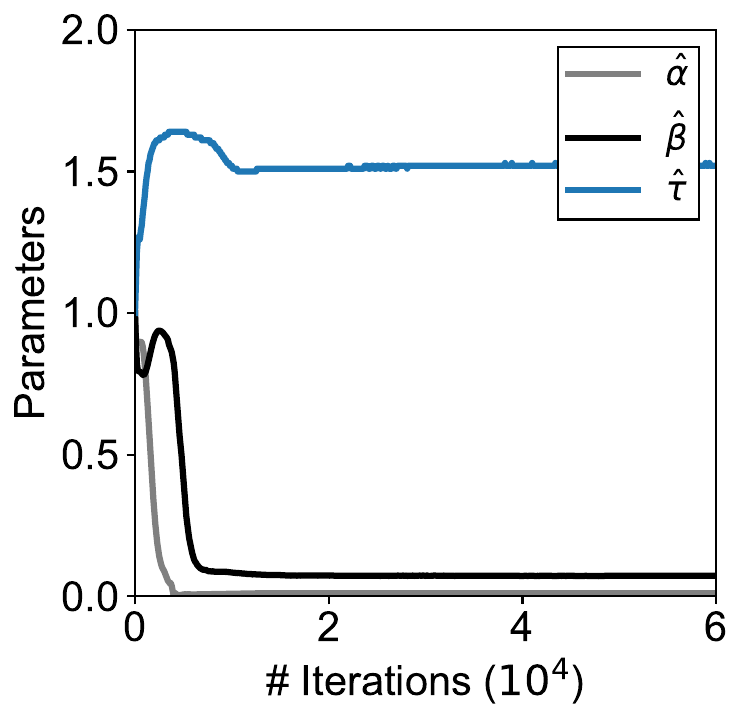}}
      \caption{CTHP parameter learning and convergence.}
  \label{Parameter convergence} 
\end{figure}

Figs \ref{Synthetic data}a--\ref{Synthetic data}b display the obtained trajectories of space-gap $p_1(t)$ (in solid grey line) and velocities $v_0(t)$ (in solid orange line) and $v_1(t)$ (in solid grey line). As can be seen, both equilibrium \eqref{eq:equil} and non-equilibrium driving conditions are considered in the synthetic dataset. It should be noted that the known set of design parameters, $\bm{\varpi}^\ast$, corresponds to a string unstable system in terms of both the $\L_2$ and $\L_\infty$ norms, check conditions \eqref{eq:L2c} and \eqref{eq:Linfc}, respectively. Thus the ACC-engaged vehicle will amplify any random perturbations of the human driven vehicle (leader).

The synthetic data were used to train the first neural network described in Section \ref{sec:PINN_Setup} towards learning the a priori known vector of CTHP parameters, $\bm{\varpi}^\ast$, of the ACC-engaged vehicle. Fig.~\ref{Synthetic_param} depicts the obtained results on the learning trajectories of $\hat{\alpha}$, $\hat{\beta}$, and $\hat{\tau}$ converging successfully to $\hat{\bm{\varpi}} = [0.0784\,\,\, 0.12\,\,\, 1.5]^\T \approx \bm{\varpi}^\ast$ after 20,000 iterations. The neural network is also calibrated to predict the entire trajectories of $p_1(t)$, $v_0(t)$, and $v_1(t)$ corresponding to the true parameter values. Fig.~\ref{Synthetic data} presents the estimated values of $\hat{p}_1(t)$, $\hat{v}_0(t)$, and $\hat{v}_1(t)$ in each point inside the entire time domain, with mean absolute errors (MAEs) of 0.0939 m,  0.1491 m/s, and 0.1509 m/s, respectively.  

This experiment underlines the ability of the developed data-driven framework and physics-inspired neural network to successfully learn the true values of the CTHP parameters and reconstruct the full range of dynamics in the space-gap and velocity profiles.

\subsection{CTHP Parameter Learning on Empirical Data}

This section demonstrates the superior predictive ability of  the proposed PiNN to learn the unknown  design parameters of stock ACC systems of different makes (see Table \ref{tab:New_Both}), using empirical data of space-gap and relative velocity from three experimental campaigns, namely Ispra-Casale, Ispra-Vicolungo, and AstaZero. It also aims to examine whether the stock ACC systems of various makes are strict string stable inside platoons using the obtained ACC parameters and the string stability criteria  \eqref{eq:L2c} and \eqref{eq:Linfc}, see Section~\ref{sec:Theoretical_Background}. 

\setlength{\tabcolsep}{5pt}
\renewcommand{\arraystretch}{1.2}
\begin{table}[tbp]
     \caption{ACC Parameter Estimation for Ispra-Casale.}\label{tab:Ispra-Casale_experiments}
     \centering
     \resizebox{\linewidth}{!}{
     \begin{tabular}{l|ccccc}
              \hline
              Experiment & \#1 & \#2 & \#3 & \#4 & \#5 \\ 
              \hline\hline
              $\hat{\alpha}$ (1/s$^2$) & 0.0104 & 0.0104 & 0.0104 & 0.0102 & 0.0103 \\
              $\hat{\beta}$ (1/s) & 0.0718 & 0.0712 & 0.0723 & 0.0709 & 0.0724 \\ 
              $\hat{\tau}$ (s) & 1.52 & 1.52 & 1.52 & 1.52 & 1.52 \\ 
              \hline
              MAE $p_1$ (m) & 0.4721 &  0.2653 & 0.4667 & 0.4519 & 0.3590 \\ 
              MAE $v_0$ (m/s) & 0.1464 & 0.1529 & 0.1925 & 0.1863 & 0.1547 \\
              MAE $v_1$ (m/s) & 0.1419 & 0.0871 & 0.1507 & 0.2183 & 0.1361 \\
              \hline
              $\mathcal{L}_2$ strict string stability & NO & NO & NO & NO & NO \\ 
              $\mathcal{L}_\infty$ strict string stability & NO & NO & NO & NO & NO \\
              \hline
          \end{tabular}}
\end{table}

\subsubsection*{Parameter Learning for Ispra-Casale}

\begin{figure*}[tbp] 
    \centering
  \subfloat[Space-gap: Empirical data, solid line; PiNN prediction, dashed line.]{%
       \includegraphics[width=.5\linewidth]{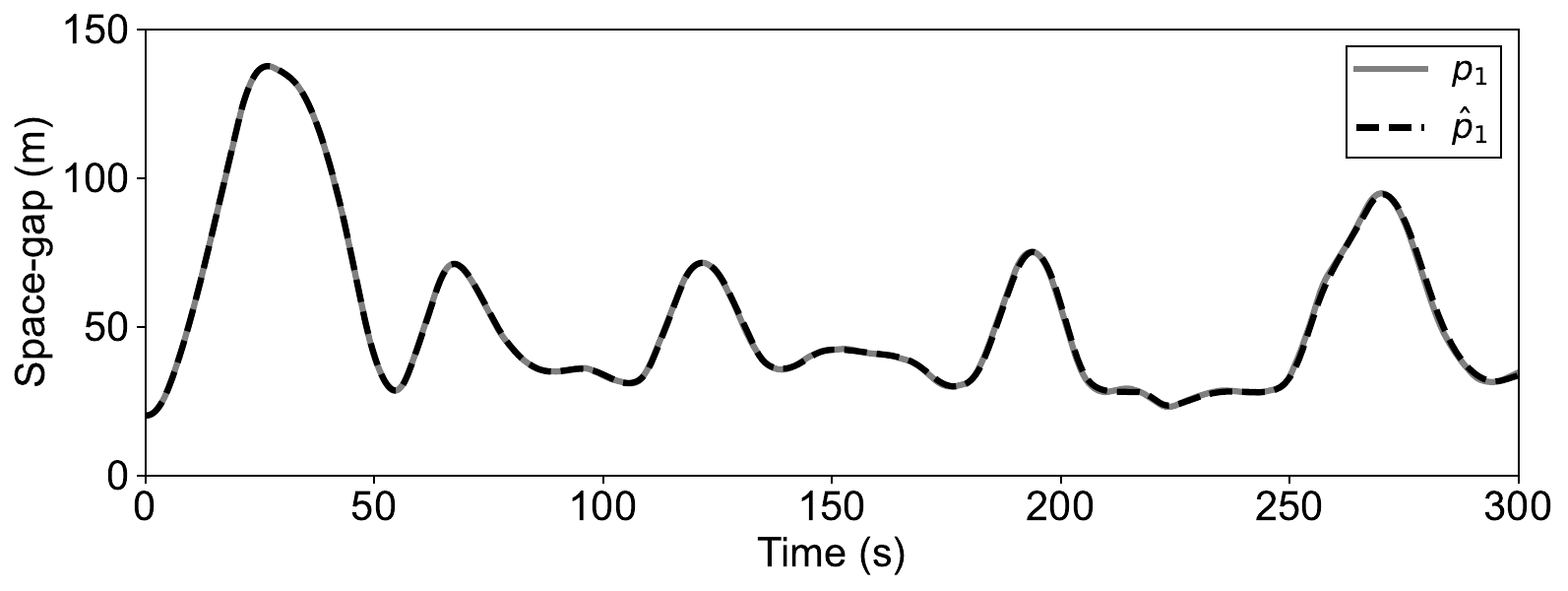}}
    \hfill 
  \subfloat[Velocity: Empirical data, solid line; PiNN prediction, dashed line.]{%
        \includegraphics[width=0.5\linewidth]{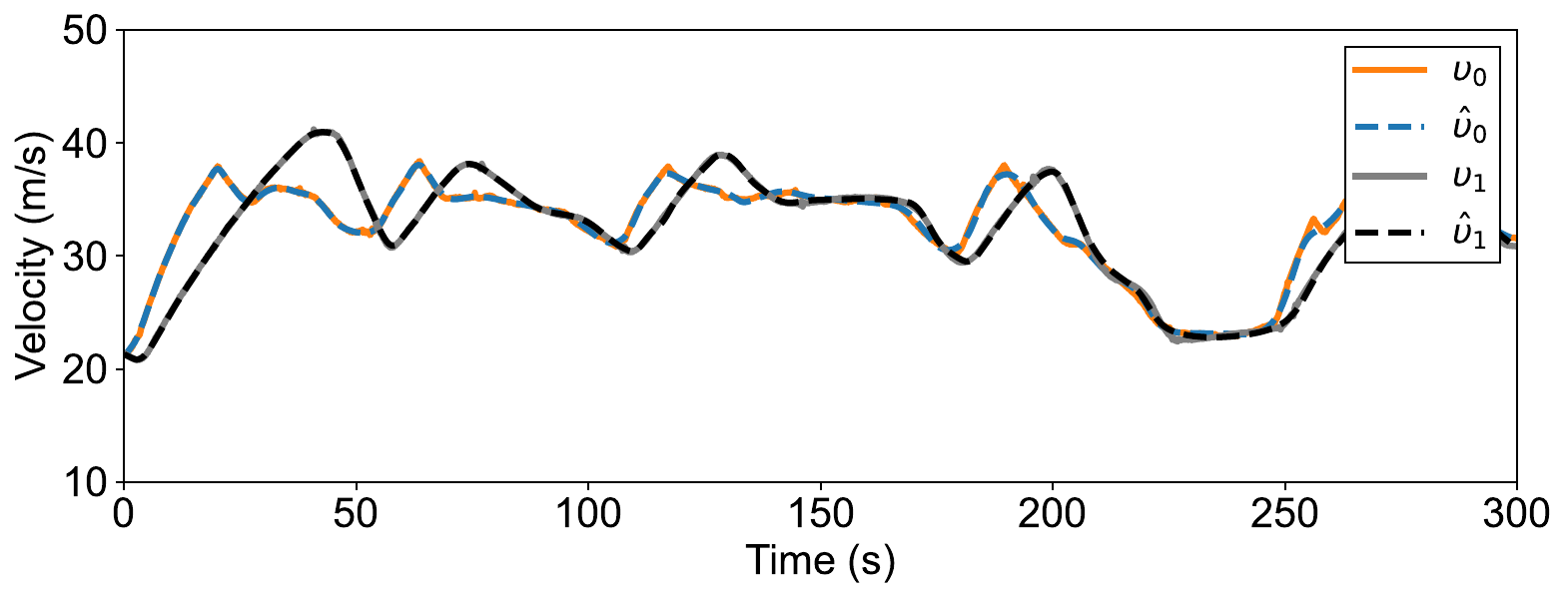}}
      \caption{Space-gap and velocity trajectories for Ispra-Casale (Exp.~\#1).}
  \label{Ispra-Casale} 
\end{figure*}

To investigate the sensitivity of parameter learning on different initial values of $\hat{\bm{\omega}}(0)$, five different training seasons of the neural network were carried out. Fig.~\ref{Ispra-Casale} displays the empirical data of space-gap and relative velocity (solid lines) used for the training. As can be seen from a careful inspection of the trajectories, the leader is engaged in various acceleration perturbations, while the follower seems to be string unstable. Table~\ref{tab:Ispra-Casale_experiments} presents the obtained results of the training for five trials with  $\hat{\alpha}$, $\hat{\beta}$ and $\hat{\tau}$ starting from different initial values. As we can see, $\hat{\alpha}$ and $\hat{\beta}$ differ to the third decimal place while $\hat{\tau}$ converges to the same value in each experiment. Using these values in the string stability criteria  \eqref{eq:L2c} and \eqref{eq:Linfc}, it turns out that the involved ACC vehicle is neither $\mathcal{L}_2$ nor $\mathcal{L}_\infty$ strict string stable. Fig.~\ref{Ispra-Casale_param} shows the parameters' convergence from the first experiment of Table~\ref{tab:Ispra-Casale_experiments}, while Fig.~\ref{Ispra-Casale} presents the estimated trajectories of space-gap and both velocities, with their MAEs being relatively low (see Table~\ref{tab:Ispra-Casale_experiments}). 

\subsubsection*{Parameter Learning  for AstaZero and Ispra-Vicolungo}

\begin{figure}[tbp] 
    \centering
  \subfloat[AstaZero]{%
       \includegraphics[width=\linewidth]{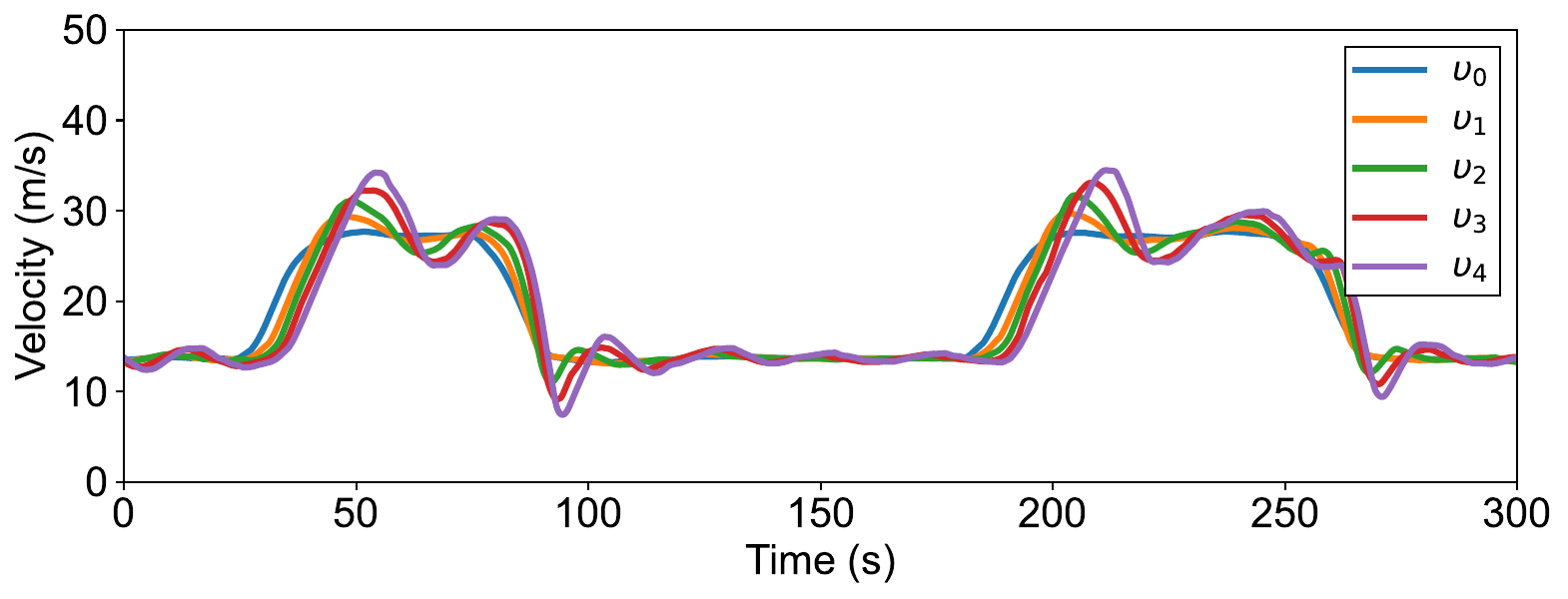}}
    \\ 
  \subfloat[Ispra-Vicolungo]{%
        \includegraphics[width=\linewidth]{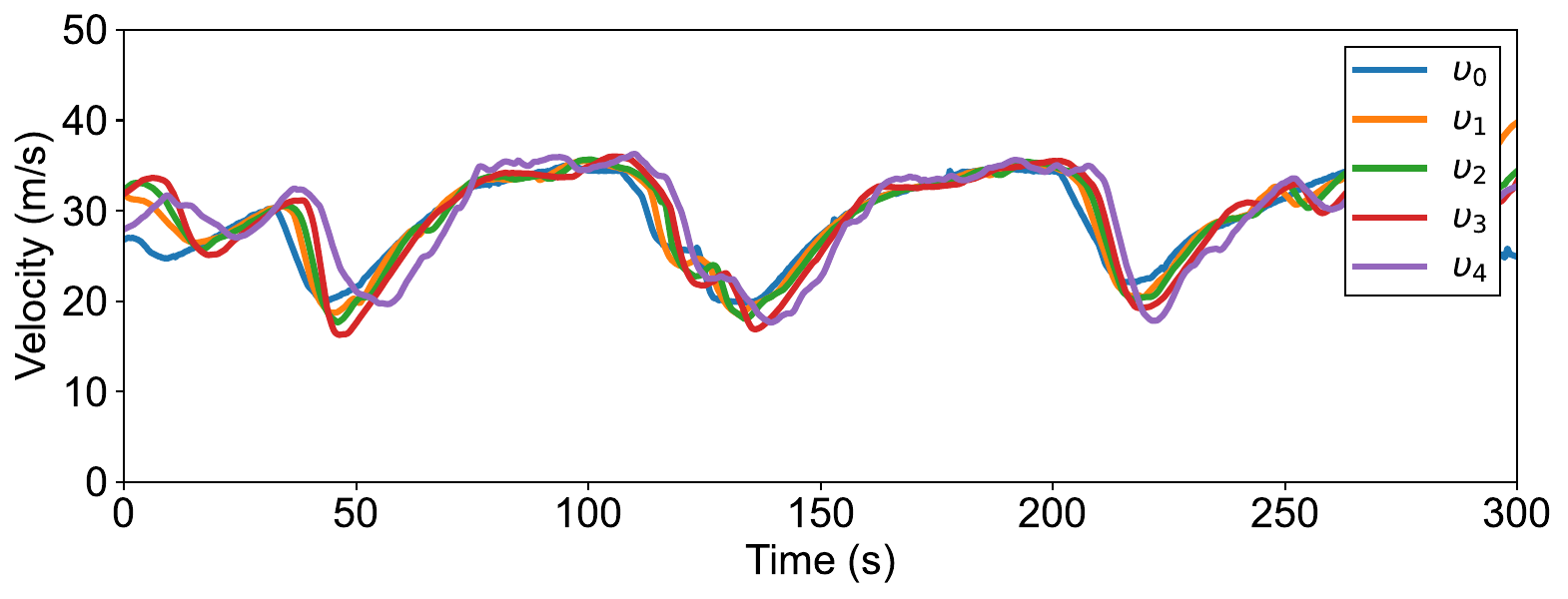}}
      \caption{Velocity profiles.}
  \label{Speed profiles} 
\end{figure}

The two experimental campaigns comprise data from real-life car-following experiments that took place in Autostrada A26 motorway between Ispra and Vicolungo and the protective environment of AstaZero test track. Each campaign involved a fleet of five ACC equipped vehicles in platoon formation, where the followers are examined under real-life traffic conditions and different perturbation events imposed by the leader, as shown in Figs~\ref{Speed profiles}a--\ref{Speed profiles}b. The data concerns the speed of each vehicle and the space-gaps between them for 300 s. 

In both campaigns, the proposed physics-inspired neural network is trained and applied to learn the CTHP parameters of the followers (vehicles indexed by $i = 1,..., 4$) and to assess the string stability of the platoons. In the sequel, two different types of platoons are considered: Case I. Homogeneous platoons where all vehicles in a platoon present the same ACC settings, i.e., $\bm{\varpi}_i \equiv \bm{\varpi} = [\alpha\,\, \beta\,\, \tau]^\T$ for all $i =1, \ldots M$; Case II. Non-homogeneous platoons where a different set of parameters $\bm{\varpi}_i = [\alpha_i\,\, \beta_i\,\, \tau_i]$, $i =1, \ldots M$, is considered for each ACC vehicle within the platoon.  

\setlength{\tabcolsep}{6pt}
\renewcommand{\arraystretch}{1.2}
\begin{table}[tbp]
     \caption{AstaZero and Ispra-Vicolungo ACC Parameter Learning Results. Case I. Homogeneous Platoons. All ACC-equipped Vehicles Present the Same CTHP Parameters.}\label{tab:Platoon_same_parameters}
     \centering
     \resizebox{\linewidth}{!}{
     \begin{tabular}{l|ccc|cc}
              \hline
               & $\hat{\alpha}$ (1/s$^2$) & $\hat{\beta}$ (1/s) & $\hat{\tau}$ (s) & $\mathcal{L}_2$ & $\mathcal{L}_\infty$ \\ 
              \hline\hline
              AstaZero & 0.0627 & 0.2630 & 1.17 & NO & NO \\
              Ispra-Vicolungo & 0.0581 & 0.3010 & 1.04 & NO & NO \\ 
              \hline
               & \multicolumn{5}{c}{AstaZero ($i = 0, \ldots,4$)}\\
              \hline\hline
              MAE $p_i$ (m) & --- & 0.3295 & \multicolumn{1}{c}{0.4731} & 0.5381 & 0.4672 \\ 
              MAE $v_i$ (m/s) & 0.1270 & 0.1436 & \multicolumn{1}{c}{0.2308} & 0.2454 & 0.3203 \\
              \hline
              & \multicolumn{5}{c}{Ispra-Vicolungo ($i = 0, \ldots,3$)}\\
              \hline\hline
              MAE $p_i$ (m) & --- & 0.4061 & \multicolumn{1}{c}{0.3271} & 0.4710 & --- \\
              MAE $v_i$ (m/s) & 0.1776 & 0.2087 & \multicolumn{1}{c}{0.2111} & 0.2563 & --- \\
              \hline
          \end{tabular}}
\end{table}

\setlength{\tabcolsep}{5pt}
\renewcommand{\arraystretch}{1.2}
\begin{table*}[tbp]
     \caption{AstaZero and Ispra-Vicolungo ACC Parameter Learning Results:\\ Case II. Non-Homogeneous Platoons, Each ACC-equipped Vehicle Present Different CTHP Parameters.}\label{tab:New_Both}
     \centering
     \resizebox{\textwidth}{!}{
     \begin{tabular}{c|c|ccc|c|cc|cc|cc}
     \hline
     Campaign & Vehicle $i$ & $\hat{\alpha}_i$ (1/s$^2$) & $\hat{\beta}_i$ (1/s) & $\hat{\tau}_i$ (s) & Mean time-gap (s) & MAE $p_i$ (m) & MAE $v_i$ (m/s) & $\mathcal{L}_2$ & $\mathcal{L}_\infty$ & Make & Model \\
     \hline\hline
     \multirow{4}{*}{AstaZero} & 0 & --- & --- & --- & --- & --- & 0.1197 & --- & --- & Audi & A8 \\
     & 1 & 0.0612 & 0.1200 & 1.19 & 1.25 & 0.2452 & 0.1492 & NO & NO & Audi & A6 \\
     & 2 & 0.1000 & 0.1470 & 1.17 & 1.23 & 0.3867 & 0.2044 & NO & NO & BMW & X5 \\
     & 3 & 0.0766 & 0.2220 & 1.16 & 1.20 & 0.5305 & 0.2585 & NO & NO & Mercedes & A-Class \\
     & 4 & 0.0409 & 0.4450 & 1.16 & 1.30 & 0.4494 & 0.3068 & NO & YES & Tesla & Model 3 \\ \hline
Platoon & Average & 0.070 & 0.234 & 1.17 & 1.25 & ---  & --- & NO & NO & --- & --- \\
     \hline
     \multirow{4}{*}{Ispra-Vicolungo} & 0 & --- & --- & --- & --- & --- & 0.1014 & --- & --- & Mitsubishi & SpaceStar \\
     & 1 & 0.0766 & 0.1660 & 1.01 & 1.03 & 0.2208 & 0.1272 & NO & NO & Ford & S-Max \\
     & 2 & 0.1760 & 0.3921 & 1.00 & 1.01 & 0.1933 & 0.1406 & NO & NO & Peugeot & 5008 \\
     & 3 & 0.0705 & 0.1930 & 1.13 & 1.18 & 0.2481 & 0.1439 & NO & NO & Kia & Niro \\
     & 4 & --- & --- & --- & 1.52 & --- & --- & --- & --- & Mini & Cooper \\ \hline
Platoon & Average  & 0.1077 & 0.2504 & 1.05 & 1.07 & --- & --- & NO & NO & --- & --- \\     \hline
   \end{tabular}}
\end{table*}

Table~\ref{tab:Platoon_same_parameters} presents the obtained CTHP parameters on both experimental campaigns for Case I. As can be seen, both campaigns are resulting in string unstable platoons with MAE between the real and estimated trajectories of space-gap and velocity being relatively low. In addition, Table~\ref{tab:New_Both} presents the collected results for both campaigns by applying one PiNN for each platoon of vehicles and considering different set of CTHP parameters, i.e., the Case II. As can be seen, the ACC design parameters  $\hat{\alpha}$, $\hat{\beta}$ and $\hat{\tau}$ of the CTHP have relatively close values across the different ACC controllers (different vehicle makes), with time-headways $\hat{\tau}$ being close enough to the true mean time-gaps; time-gap is specified as the fraction of the space-gap between two vehicles to the speed of the following vehicle. Comparing with Table~\ref{tab:Platoon_same_parameters} (notice the averages over the platoon in Table~\ref{tab:New_Both}) the CTHP parameter values in Case II are seen to be consistent with the values obtained in Case I, which underlines the superiority of the proposed physics-constrained and data-informed neural network to learn the ACC model parameters with different assumptions and architectures for two real-life car-following campaigns. Table~\ref{tab:New_Both} also presents the MAEs between the estimated trajectories (regarding space-gap and velocity) and the real ones. The obtained MAEs across the different PiNN's implementations are consistent to previous works on the parameter identification of commercial ACC systems \cite{Knoop:2019,Wang:2021,Gunter:2021,Ampountolas:2023}. 

\begin{figure}[tbp] 
   \centering
      \includegraphics[width=\linewidth]{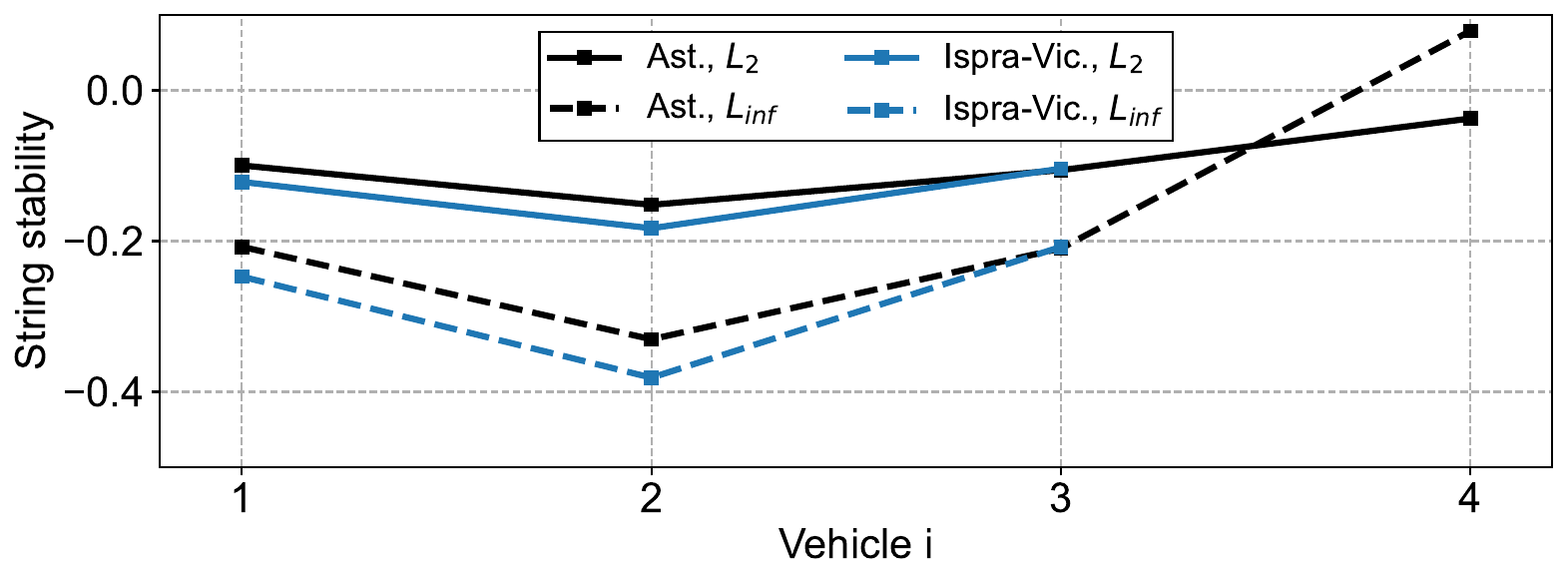}
      \caption{$\L_2$ and $\L_\infty$ string stability evolution over the platoon in each campaign.}
 \label{Stability} 
\end{figure}

\subsection{String Stability and Bode Plots}

This section discusses the string stability of the calibrated CTHP models for both experimental campaigns. First, the $\mathcal{L}_2$ and $\mathcal{L}_\infty$  string stability conditions (see \eqref{eq:L2c} and \eqref{eq:Linfc}) are calculated and checked. Table~\ref{tab:New_Both} displays the obtained results, see column two from right. As can be seen, the ACC followers are neither $\mathcal{L}_2$ nor $\mathcal{L}_\infty$ strict string stable, except the last vehicle in AstaZero, which is seen to be $\mathcal{L}_\infty$ string stable. This is possible since $\L_2$ stability is stronger than the $\mathcal{\L}_\infty$ stability, see condition \eqref{eq:L2Linf}. Also, the last follower ($i = 4$) in the Ispra-Vicolungo campaign was driving manual, and hence, there are no findings. Finally, Fig.~\ref{Stability} depicts the evolution of $\mathcal{L}_2$ and $\mathcal{L}_\infty$ values of the ACC followers inside the platoon. A valley is formed in the middle of the platoon with $\mathcal{L}_2$ values being lower than those of $\mathcal{L}_\infty$, in both campaigns. The observed behavior of the two stability criteria suggests that the leader's ($i = 0$) perturbation in both trials is upstream propagated on the ACC vehicles with $i = 1, 2$ and then somehow dissipated by the ACC vehicles with $i = 3, 4$.    

\begin{figure*}[tbp] 
    \centering
  \subfloat[AstaZero]{%
       \includegraphics[width=\columnwidth]{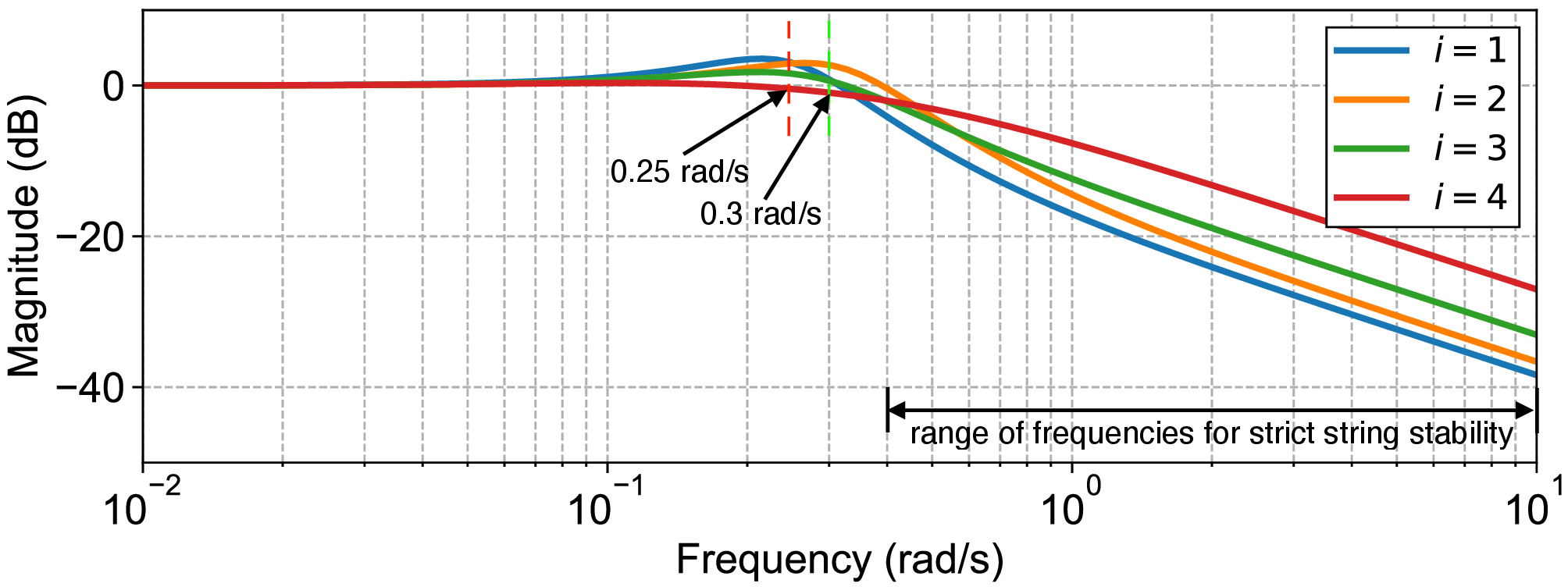}}\hfill
  \subfloat[Ispra-Vicolungo]{%
        \includegraphics[width=\columnwidth]{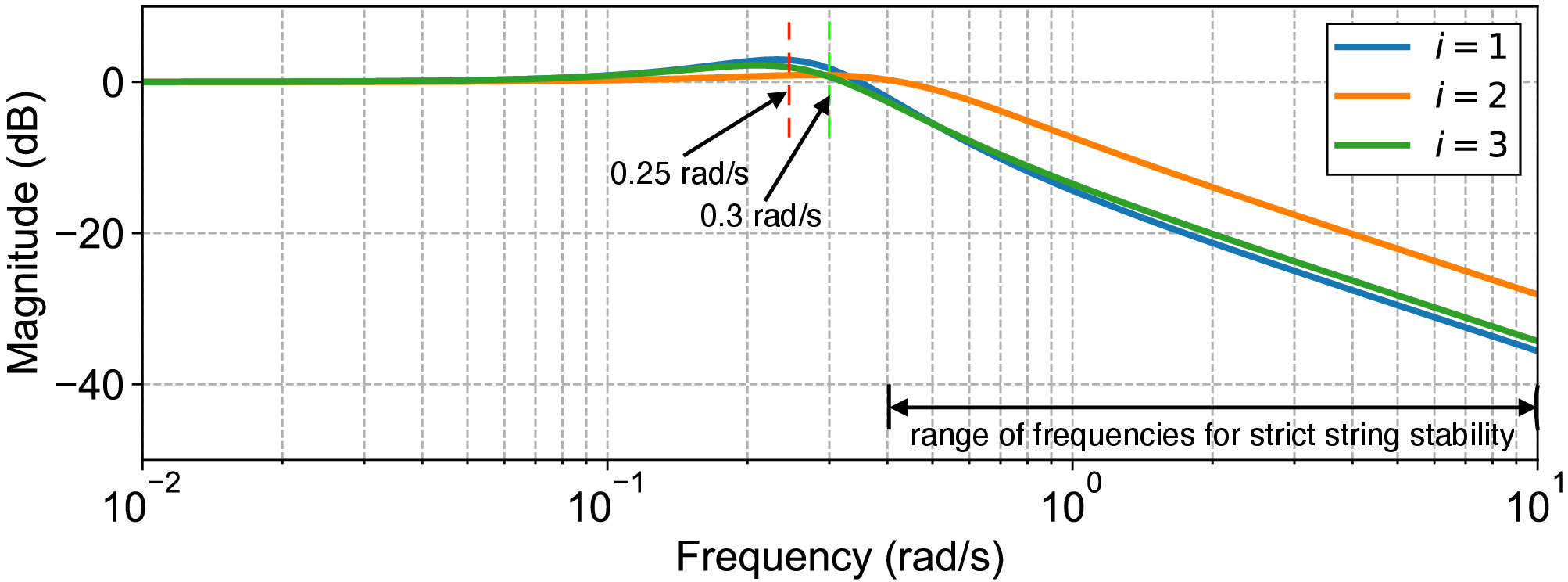}}
      \caption{Bode plots for the ACC vehicles involved in AstaZero and Ispra-Vicolungo campaigns, see Table \ref{tab:New_Both}. Both Bode plots indicate positive amplitude at a given frequency  (i.e., perturbation) led to string unstable platoons. In the range of frequencies $[0.4,1]$ rad/s all calibrated ACC models will dissipate disturbances (i.e., being string stable). (a) AstaZero: At $\omega = 0.25$ rad/s vehicle $i=4$ is \emph{marginal string stable} while all the other vehicles are \emph{string unstable}. On the other hand at $\omega = 0.3$ rad/s vehicle $i=4$ is \emph{string stable}, $i=2$ remains string unstable and $i=1, 3$ are \emph{marginal string stable}. (b) Ispra-Vicolungo: Same observations for different vehicles.}
  \label{Bode_plots} 
\end{figure*}

To further interpret the obtained string stability results, Fig.~\ref{Bode_plots} presents the Bode plots of the calibrated CTHP models for both campaigns. The Bode plot is developed using the velocity-to-velocity transfer function \eqref{eq:tf_CTHP} and the calibrated CTHP model parameters for each vehicle in the platoon (see Table \ref{tab:New_Both}). Provided Definition \ref{defn:String_Stability} and  string stability condition \eqref{eq:L2_f}, a positive amplitude at the Bode plot at a given frequency (i.e., disturbance) indicates that the disturbance will propagate through the platoon; while a negative amplitude indicates that the disturbance will dissipate. For zero amplitude the platoon is marginal string stable (or weak string stable). 

As can be seen in both campaigns (see Fig.~\ref{Bode_plots}), the ACC-engaged vehicles (vehicles $i = 1, 2, 3, 4$ in AstaZero and $i = 1, 2, 3$ in Ispra-Vicolungo) could not compensate instabilities generated by the leader
(vehicle $i = 0$). However there is a range of frequencies (in [0.4,1] rad/s) over which the calibrated ACC models in both campaigns will dissipate disturbances. Precisely, disturbances with frequencies less than 0.4 rad/s are amplified, while larger frequency disturbances will be dissipated along the platoon. The largest amplitude ($\approx1.2$ dB) for both campaigns occurs at $\omega = 0.25$ rad/s. These observations suggest that a stronger condition than $\mathcal{L}_2$ strict string stability would be to force a sharper decrease of the Bode plot for low frequencies. Note that this is challenging since the displayed Bode plots correspond to a second order system with a zero at the left-half complex plane (see transfer function \eqref{eq:tf_CTHP}).

\begin{figure*}[tbp]\centering 
\begin{tabular}{cc}        \includegraphics[width=\columnwidth]{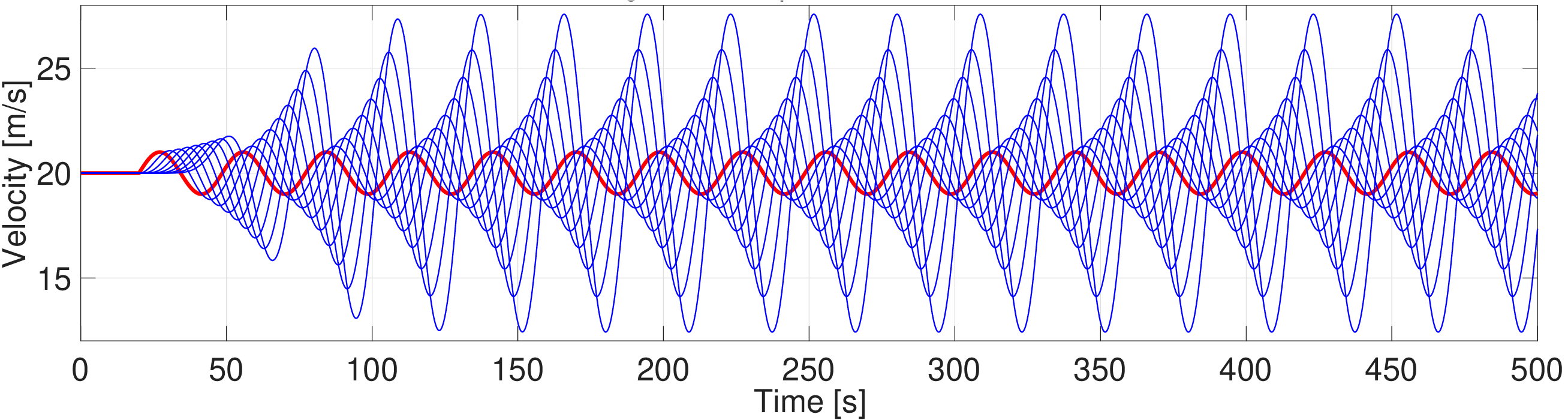}
     & \includegraphics[width=\columnwidth]{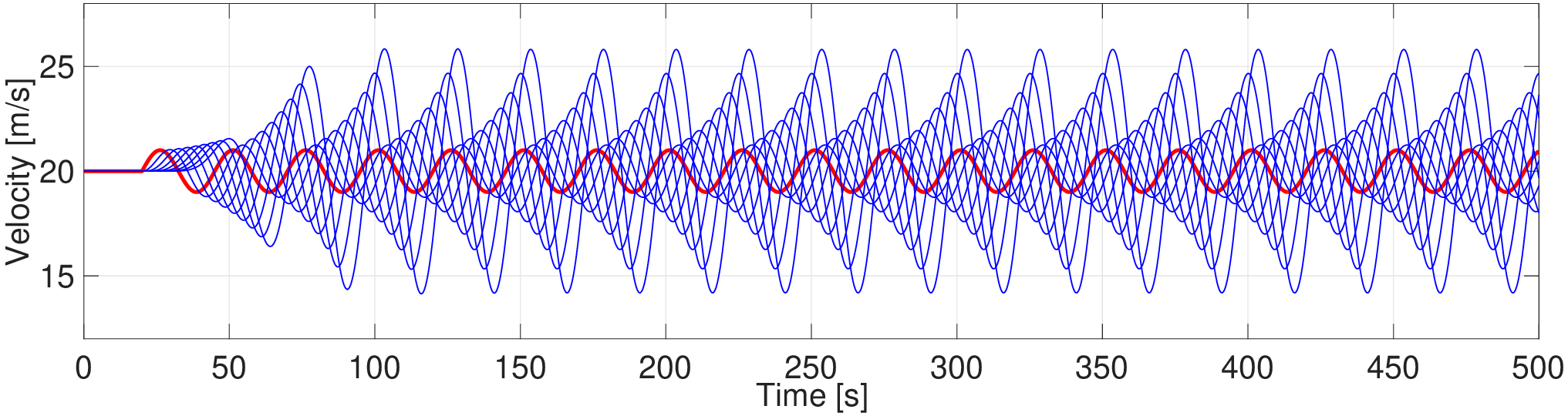}\\
     \small (a) String unstable: Sinusoidal perturbation at $0.25$ rad/s & 
     \small (c) String unstable: Sinusoidal perturbation at $0.3$ rad/s\\[4pt]
\includegraphics[width=\columnwidth]{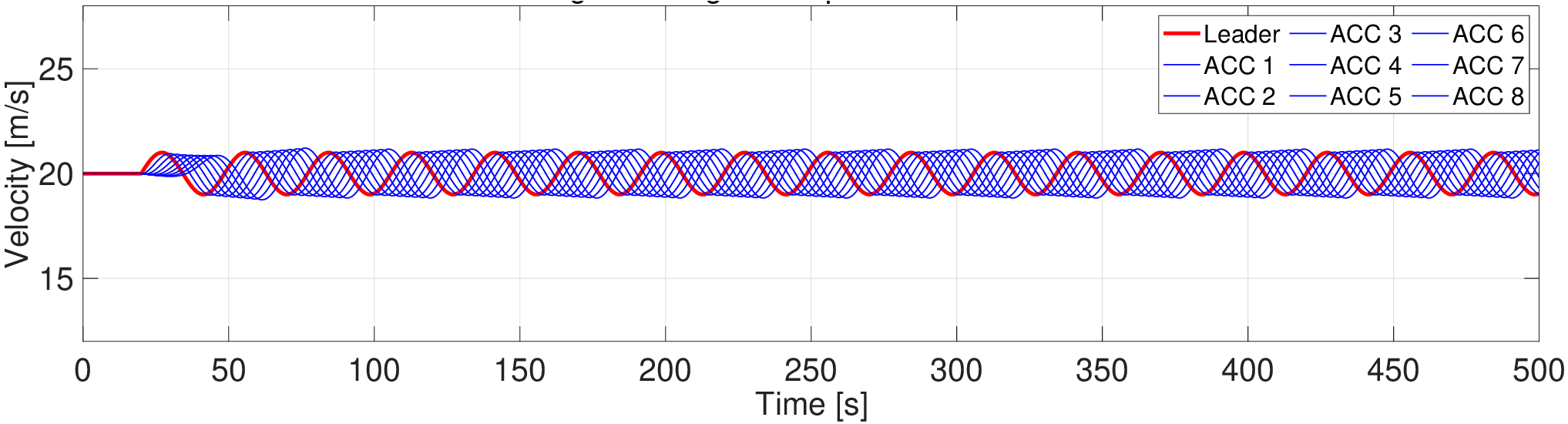}
     & \includegraphics[width=\columnwidth]{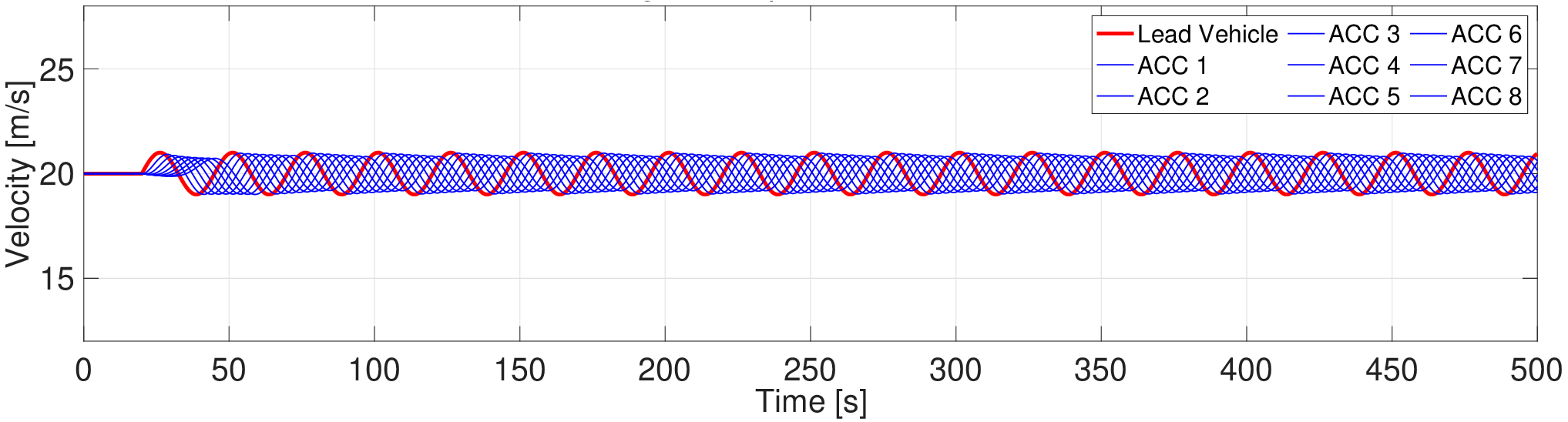}\\
     \small (b) Marginal string stable: Sinusoidal perturbation at $0.25$ rad/s& 
     \small (d) String stable: Sinusoidal perturbation at $0.3$ rad/s
\end{tabular}   
      \caption{Velocity profiles for a simulation of eight (8) ACC-equipped vehicles following a lead vehicle performing a sinusoidal perturbation with a magnitude of 1 m/s at two different frequencies of the Bode plot. The ACC vehicles in all experiments obey string unstable CTHP model parameters (see \eqref{eq:L2c} and \eqref{eq:Linfc}).}\label{fig:Sim_8Vehs} 
\end{figure*}

To illustrate that disturbances for some frequencies remain bounded or dissipated even with a string unstable ACC system (i.e., CTHP model parameters verified as string unstable by \eqref{eq:L2c} and \eqref{eq:Linfc}), Fig.~\ref{fig:Sim_8Vehs} depicts the velocity profiles of eight (8) ACC-equipped vehicles following a lead vehicle performing a sinusoidal perturbation for a simulation of $500$ s. The lead vehicle (shown in solid red colour) drives at 20 m/s for 20 s, while all following ACC-engaged vehicles initialized at the corresponding equilibrium velocity and space-gap ($\propto \tau v$). Then the lead vehicle performs a sinusoidal perturbation with a magnitude of 1 m/s at $\omega = 0.25$ rad/s and $\omega = 0.3$  rad/s, respectively. The ACC-engaged vehicles adopting the CTHP with calibrated parameters corresponding to the string unstable vehicles $i=3$ and $i=4$ from the AstaZero campaign (see Table \ref{tab:New_Both} and Fig.~\ref{Bode_plots}a).  As can be seen in Fig.~\ref{fig:Sim_8Vehs}a and Fig.~\ref{fig:Sim_8Vehs}c, the ACC vehicles amplify the disturbance along the platoon with calibrated CTHP parameters corresponding to vehicle $i=3$, while dissipate  the oscillation with calibrated parameters for $i=4$ (see Fig.~\ref{fig:Sim_8Vehs}b and Fig.~\ref{fig:Sim_8Vehs}d).

\section{Conclusions and Outlook}\label{sec:Conclusions}

The parameter learning of commercially implemented ACC systems is challenging, since the core functionality of those systems (proprietary control logic and its parameters) is not publicly available. This work unveiled that physics-constrained and data-informed neural networks can be used as a surrogate  model to capture the ACC vehicle longitudinal dynamics and efficiently infer the unknown parameters of the constant time-headway policy, often deployed in stock ACC systems of various makes in automotive industry.    

The findings of this paper demonstrate the ease in which PiNNs perform in learning the unknown ACC design parameters. More specifically, PiNNs may retrieve successfully the true ACC parameters, based on synthetic data, as well as to deliver estimates of the unknown design parameters of stock ACC systems, based on empirical observations of space-gap and relative velocity. This is confirmed by the similar ACC parameter values found among the different ACC controllers (different vehicle makes) in the three experimental campaigns. Applying the string stability criteria to the obtained results showed that the stock ACC systems of various makes tend to be string unstable inside the platoon. This result highlights that further research is needed to achieve string stable platoons with a positive effect on traffic flow, capacity and throughput \cite{Apostolakis_Energy:2023}. This will allow the large-scale deployment of ACC-engaged vehicles on freeways.

Despite the ease in which PiNNs successfully learn the unknown ACC design parameters are subject to limitations: 
\begin{itemize}
    \item The parameter learning problem  is data-driven, necessitating the availability of car-following experimental data with ACC vehicles involved to train the PiNN. 
    \item The time domain is restricted to 300 s due to the computation effort required to train larger neural networks.
    \item The computation effort is higher compared to rival methods (e.g., Kalman filter \cite{Ampountolas:2023}), although the training can be performed offline. However, once the model is trained, the prediction of future trajectories is straightforward.
\end{itemize}


\bibliographystyle{IEEEtran}
\bibliography{IEEEabrv,New_Bibliography}

%

\begin{IEEEbiography}[{\includegraphics[width=1in,height=1.25in,clip,keepaspectratio]{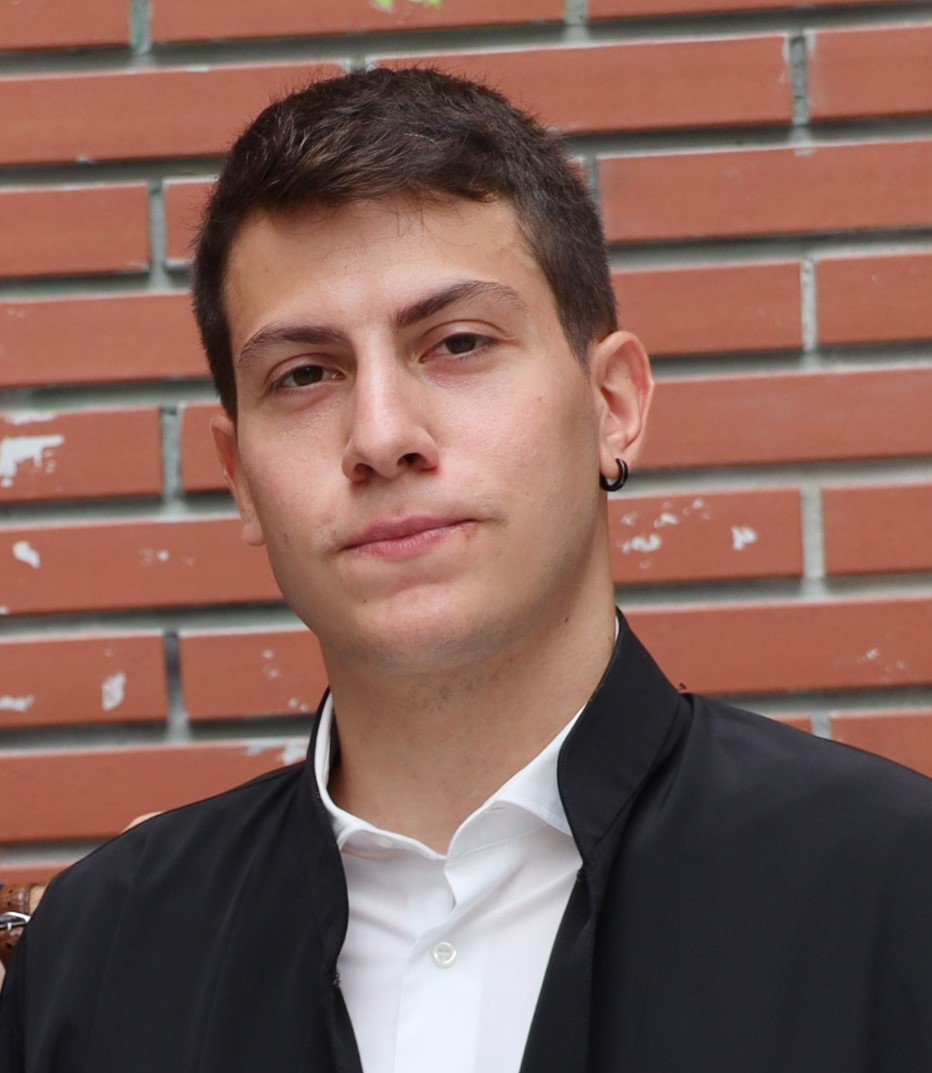}}]{Theocharis Apostolakis} received the Dipl.Ing.~degree in Mechanical Engineering from the University of Thessaly, Greece, in 2022. He is currently a PhD candidate with the Automatic Control and Autonomous Systems Laboratory at the same department. His research interests include control systems and optimization.
\end{IEEEbiography}

\vfill

\begin{IEEEbiography}[{\includegraphics[width=1in,height=1.25in,clip,keepaspectratio]{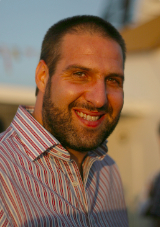}}]{Konstantinos Ampountolas}
(Member, IEEE) received the Dipl.Ing.\ degree in production engineering and management, the M.Sc.\ degree in operations research, and the Ph.D.~degree in engineering from the Technical University of Crete, Greece, in 1999, 2002, and 2009, respectively. He was a Senior Lecturer with the James Watt School of Engineering, University of Glasgow, U.K., from 2013 to 2019, a Research Fellow with the {\' E}cole Polytechnique F{\' e}d{\' e}rale de Lausanne (EPFL), Switzerland, from 2012 to 2013, a Visiting Researcher Scholar with the University of California at Berkeley, Berkeley, CA, USA, in 2011, and a Post-Doctoral Researcher with the Centre for Research \& Technology Hellas, Greece, in 2010. He was also a short-term Visiting Professor with the Technion–Israel Institute of Technology, Israel, in 2014, and the Federal University of Santa Catarina, Florianópolis, Brazil, in 2016 and 2019. Since 2019, he has been an Associate Professor with the Department of Mechanical Engineering, University of Thessaly, Greece. His research interests include automatic control and optimization and their applications to transport networks and systems. Dr.~Ampountolas has served as the Editor for \emph{Transportation} of \emph{Data in Brief}  from 2018 to 2019, as an Associate Editor for the \emph{Data Science for Transportation}  from 2018 to 2020, and on the editorial advisory boards of \emph{Transportation Research Part C: Emerging Technologies} from 2014 to 2021, and \emph{Transportation Research Procedia} since 2014.
\end{IEEEbiography}

 \vfill
\end{document}